\documentclass[
reprint,
amsmath,amssymb,
pra
]{revtex4-2}

\usepackage{graphicx}
\graphicspath{{figures/}} 
\usepackage{dcolumn}
\usepackage{bm}
\usepackage{physics}
\usepackage{braket}
\usepackage{placeins}
\usepackage[colorlinks=true]{hyperref}
\usepackage{upgreek}
\usepackage{multirow}
\usepackage{dsfont}

\begin{document}

\let\oldaddcontentsline\addcontentsline
\renewcommand{\addcontentsline}[3]{}

\title{
Individually-addressed quantum gate interactions using dynamical decoupling\\
}

\begin{abstract}
    A leading approach to implementing small-scale quantum computers has been to use laser beams, focused to micron spot sizes, to address and entangle trapped ions in a linear crystal.
    Here we propose a method to implement individually-addressed entangling gate interactions, but driven by microwave fields, with a spatial-resolution of a few microns, corresponding to $10^{-5}$ microwave wavelengths.
    We experimentally demonstrate the ability to suppress the effect of the state-dependent force using a single ion, and find the required interaction introduces $3.7(4)\times 10^{-4}$ error per emulated gate in a single-qubit benchmarking sequence.
    We model the scheme for a 17-qubit ion crystal, and find that any pair of ions should be addressable with an average crosstalk error of $\sim 10^{-5}$.
 \end{abstract}

\author{M.\,C.\,Smith}
\thanks{These authors contributed equally.}
\affiliation{Clarendon Laboratory, Department of Physics, University of Oxford, Parks Road, Oxford OX1 3PU, U.K.}

\author{A.\,D.\,Leu}
\thanks{These authors contributed equally.}
\affiliation{Clarendon Laboratory, Department of Physics, University of Oxford, Parks Road, Oxford OX1 3PU, U.K.}

\author{M.\,F.\,Gely}
\thanks{These authors contributed equally.}
\affiliation{Clarendon Laboratory, Department of Physics, University of Oxford, Parks Road, Oxford OX1 3PU, U.K.}

\author{D.\,M.\,Lucas}
\affiliation{Clarendon Laboratory, Department of Physics, University of Oxford, Parks Road, Oxford OX1 3PU, U.K.}

\date{\today}

\maketitle

\begin{figure}
      \centering
      \includegraphics[width=0.45\textwidth]{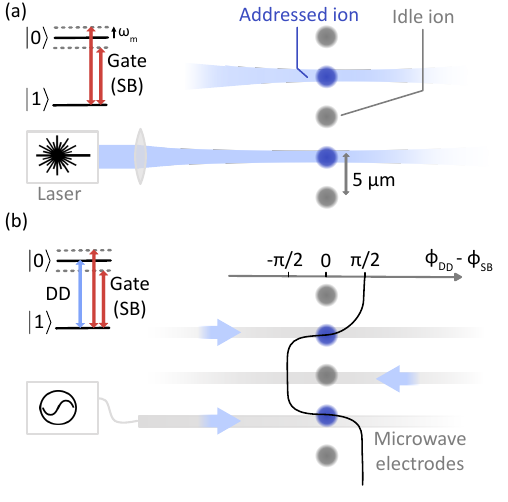}
      \caption{
      \textbf{Two-qubit gate addressing through spatially varying dynamical decoupling.}
      \textbf{(a)} A laser is commonly used to address ions by focusing beams to small spot sizes \cite{nagerl1999}.
      To drive the two-qubit gate interaction -- state-dependent force (SDF) -- two sideband (SB) laser tones are applied, symmetrically detuned from the qubit frequency by the motional frequency $\omega_m$.
      \textbf{(b)} An SDF can also be driven using microwave currents, tuned to SB frequencies, and injected through electrodes (light grey) passing under the ions.
      Due to the long wavelength of microwaves, this force will affect all ions trapped within the same potential well.
      Our proposed method for targeting a single ion requires an additional dynamical decoupling tone (DD) which drives the qubit resonantly.
      By spatially varying the phase difference between DD and SB drives -- through interference of the field generated by the different electrodes -- the effect of the SDF is either enabled (DD and SB in-phase) or suppressed (DD and SB in quadrature).
      }
      \label{fig:1}
  \end{figure}

  Dynamical decoupling (DD)~\cite{viola1998, viola1999, hahn1950, carr1954, meiboom1958} is commonly used across many quantum computing platforms to reduce the loss of quantum information caused by decoherence~\cite{suter2016, ezzell2023dynamical, Biercuk2009, Damodarakurup2009,Jiangfeng2009, jenista2009, wang2012}.
  DD aims to suppress the interaction between a quantum system and its environment by imposing a time-dependence on the interaction, and averaging out the net effect on the system.
  This technique is used for enhancing quantum memories~\cite{ezzell2023dynamical,Biercuk2009,Damodarakurup2009,Damodarakurup2009,Jiangfeng2009,jenista2009,wang2012} and reducing errors during logical gate operations~\cite{tan2013,harty2016,manivitz2017,barthel2022robust}.
  Recently, it has been put forth as a promising method to reduce crosstalk from coherent control pulses, for spin~\cite{Buterakos2018} and superconducting~\cite{Tripathi2022} qubits.
  In this article, we extend this idea to selectively decouple qubits within the same register from gate interactions, enabling individual qubit addressing.
  We demonstrate this idea using trapped ion qubits driven by near-field microwaves~\cite{ospelkaus2008,ospelkaus2011,harty2014,harty2016,zarantonello2019,Hahn2019a,Weidt2016,srinivas2021}.
  More specifically, we show how DD can enable individually-addressed $\hat\sigma_x\otimes\hat\sigma_x$ gate interactions~\cite{ospelkaus2008,ospelkaus2011,harty2014,harty2016,zarantonello2019,Hahn2019a}, addressing a key challenge with microwave-driven trapped ions over the laser-driven alternative.
  Microwave technology offers attractive features for ion trap scalability: robustness, cost and size, straightforward amplitude and phase control, and easy integration of waveguides onto surface traps.
  However, whilst laser beams can be focused onto individual ions~\cite{nagerl1999}, the centimeter-wavelength of microwaves requires alternate techniques to address ions confined to the same potential well~\cite{piltz2014,warring2013,randell2015,craik2017,sutherland2022,srinivas2022,leu2023}.
  Individually-addressed microwave-driven two-qubit gates have previously been demonstrated using magnetic gradient induced coupling (MAGIC)~\cite{Khromova2012}, however this technique cannot be used on magnetic field insensitive ``clock'' qubits and is therefore susceptible to magnetic field fluctuations.
  We first show how DD can be used to suppress the effect of the interaction which drives two-qubit gates, the state-dependent force (SDF) ~\cite{sorenson2000, sorenson1999, milburn2000}.
  The error associated with the suppression of this ``state-dependent displacement'' (SDD) is measured to be much lower than typical two-qubit gate errors: $3.7(4)\times 10^{-4}$.
  Secondly, we implement a spatially varying DD phase to selectively suppress or enable the SDD with $\sim 1~\upmu$m spatial resolution, as schematically illustrated in Fig.~\ref{fig:1}.
  We extrapolate our results to a larger register of ions, showing how this technique could enable all-to-all connectivity in a chain of ions solely by varying the amplitude of microwave currents in three electrodes.

  For this demonstration, a single ion is displaced to different positions to emulate different ions in a chain.
  Our experiments are carried out at room temperature on a segmented electrode surface ``chip'' trap characterised in Ref.~\cite{weber2022cryogenic}.
  The trap features an on-chip microwave (MW) resonator with a single ion trapped $40~\upmu\text{m}$ above the chip surface.
  Our qubit is defined by the hyperfine states $|1\rangle = |F=4, M_{F}=1 \rangle$ and $|0\rangle =|F=3, M_{F}=1 \rangle$ of the ground state manifold 4$^{2}$S$_{1/2}$ of $^{43}$Ca$^{+}$, which form a clock qubit at our static magnetic field strength of 28.8 mT.
  These states are connected by a magnetic dipole transition with frequency splitting $\omega_{q} = 2\pi\times 3.1$ GHz.
  To drive this transition, we use near-field radiation generated by the currents propagating in the MW resonator.
  The field gradient is used to generate the SDF enabling two-qubit gates.

  \begin{figure}
      \centering
      \includegraphics[width=0.45\textwidth]{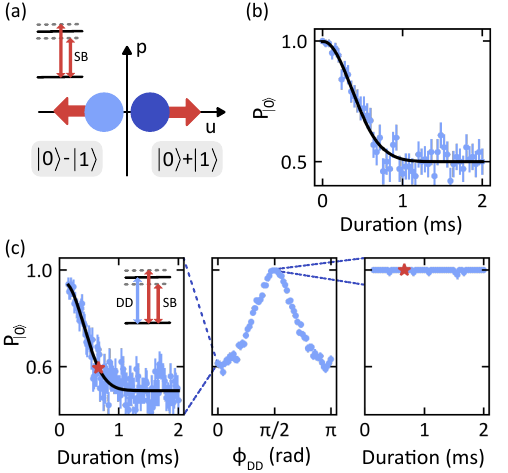}
      \caption{
          \textbf{Suppressing the effect of the state-dependent force through dynamical decoupling.}
          \textbf{(a)} Position-momentum (u-p) diagram illustrating the effect of the state-dependent force (SDF) on the ion's motional state.
      Starting with qubit state $\ket{0}$, a measurement in the qubit basis ($\ket{0}$, $\ket{1}$), would show the qubit evolving from state $\ket{0}$ to $(\ket{0}\pm\ket{1})/\sqrt{2}$.
      \textbf{(b)} Measured (blue) and fitted (black) probability $P_{\ket{0}}$ of measuring the initial state $\ket{0}$ after driving an SDF using two microwave sideband (SB) tones.
      \textbf{(c)} To inhibit motional entanglement, a dynamical decoupling (DD) tone drives the qubit with a variable phase $\phi_\text{DD}$ (relative to the average phase of the two sidebands).
      We measure $P_{\ket{0}}$ for a fixed pulse duration (red star), such that suppression of the state-dependent displacement (SDD) corresponds to $P_{\ket{0}}>0.6$.
      Measurements were taken at a fixed ion position (null of the RF trapping field).
      The SDD is unaffected ($P_{\ket{0}}\approx 0.6$) or suppressed ($P_{\ket{0}}\approx 1$) for $\phi_\text{DD} = 0$ and $\phi_\text{DD} = \pi/2$ respectively.
      All error bars indicate 68$\%$ confidence intervals.
      }
      \label{fig:2}
  \end{figure}

  An SDF is generated by driving the red and blue motional sidebands of the qubit transition at frequencies $\omega_q\pm\omega_m$ (see Sec. S1) where $\omega_m$ corresponds to the frequency of one of the motional modes of the ion -- here we use the in-plane radial mode ($\omega_m = 2 \pi \times 5.77$ MHz).
  The SDF displaces the state of this mode in position-momentum phase space where the sign of the displacement depends on the qubit state \cite{Wineland1996CatState}.
  By slightly detuning from the red and blue sidebands, the motion describes loops in phase space -- the central mechanism of M\o{}lmer-S\o{}rensen (MS) two-qubit gates \cite{sorenson2000, sorenson1999, milburn2000} -- but we will mostly focus on the resonant case for simplicity, and without loss of applicability to the detuned case.
  Under a rotating wave approximation and ignoring negligible off-resonant carrier driving, the SDF is described by the Hamiltonian (see Sec. S1)
  \begin{equation} \label{eq:MW_Hamiltonian_sdf}
      \hat H_{\text{SDF}} = \frac{\hbar}{2} \Omega_{\text{SB}} \hat{\sigma}_{x} \left( \hat{a} + \hat{a}^{\dagger} \right)\ ,
  \end{equation}
  here written in the interaction frame with respect to the qubit and motion, where $\Omega_{\text{SB}}$ is the sideband interaction strength, proportional to the gradient of the field driving the ion.
  Re-written as
  \begin{equation} \label{eq:MW_Hamiltonian_sdf2}
      \hat H_{\text{SDF}} =  \frac{\hbar}{2} \Omega_{\text{SB}} \ket{+}\bra{+} \left( \hat{a} + \hat{a}^{\dagger} \right) - \frac{\hbar}{2} \Omega_{\text{SB}} \ket{-}\bra{-} \left( \hat{a} + \hat{a}^{\dagger} \right)
  \end{equation}
  the state-dependent force is made explicit: the motion experiences a force which is positive if the qubit is in state $\ket{+}=(\ket{0}+\ket{1})/\sqrt{2}$ and negative if the qubit state is $\ket{-}=(\ket{0}-\ket{1})/\sqrt{2}$.
  We propose to selectively suppress the state-dependent displacement (SDD) driven by the SDF through the choice of the phase of an additional dynamical decoupling (DD) tone driving the qubit transition.
  For the MW electrode geometry used in this experiment, the dynamical decoupling (DD) drive phase changes with ion position, localising the effect to a chosen region of space.
  The DD tone drives the interaction
  \begin{equation}
      \hat H_{\text{DD}}  = \frac{\hbar}{2} \Omega_{\text{DD}} e^{i \phi_{\text{DD}}} \hat{\sigma}_{+}+ \text{h.c.}
      \label{eq5:equation_name}
  \end{equation}
  where $\hat{\sigma}_{+} = |1\rangle\langle0|$, $\Omega_{\text{DD}}$ is the strength of the DD drive, and the phase $\phi_{\text{DD}}$ depends on the ion position and the phase of the injected microwave current.
  Note that if the DD and SDF are in-phase, $\phi_{\text{DD}}=k\pi$ ($k\in  	\mathbb{Z}$), the Hamiltonians commute, $[\hat H_{\text{SDF}}, \hat H_{\text{DD}}] = 0$, and the dynamical decoupling does not alter the SDF dynamics.
  Whilst this has already been demonstrated in two-qubit gates~\cite{harty2016}, here we also make use of the in quadrature case.

  When the DD and SDF are in quadrature $\phi_{\text{DD}}=\pi/2+k\pi$, DD driving can suppress the effect of the qubit-motion interaction.
  The DD tone will drive Rabi oscillations between the states on which the SDF depends $\ket{+}\leftrightarrow\ket{-}$, and, if $ \Omega_{\text{DD}}\gg \Omega_{\text{SB}}$, the rapid changes of SDF direction will stop the motion from gaining significant amplitude in phase space.
  We propose to combine this technique with a microwave electrode geometry that enforces $\phi_{\text{DD}}=k\pi$ for ion positions where the SDF is desired, and $\phi_{\text{DD}}=\pi/2+k\pi$ for ion positions where the SDF is undesired.
  First, however, we experimentally demonstrate SDD suppression at a fixed ion position.

  Choosing to resonantly drive the sidebands, and subsequently measuring the qubit state, provides a straightforward measurement of the SDD and its suppression.
  We first demonstrate the effect of the SDF with no DD driving, as shown in Fig.~\ref{fig:2}(a).
  Each experimental cycle starts by preparing the state $\ket{0}\ket{0_m}$, where $\ket{0_m}$ designates the motional ground state through microwave-enhanced optical pumping~\cite{harty2014} and Raman-laser driven sideband cooling~\cite{weber2022cryogenic}.
  A small thermal population does remain in the motional mode (see Sec. S3), which is included in all our simulations, but omitted here for simplicity.
  After driving the motional sidebands for a duration $t$, the initial state
  \begin{equation}
      \ket{0}\ket{0_m}= \frac{1}{\sqrt{2}} \left( \ket{+}+\ket{-} \right)\ket{0_m}
  \end{equation}
   will evolve under the SDF to
  \begin{equation} \label{eq4:equation_name}
  \begin{aligned}
      \ket{\psi}&=\frac{1}{\sqrt{2}} \left( \ket{+}\ket{+\alpha}+\ket{-}\ket{-\alpha} \right)\\
      &=\frac{1}{2} \ket{0} \left(\ket{\alpha} + \ket{-\alpha} \right) + \frac{1}{2}\ket{1} \left(\ket{\alpha} - \ket{-\alpha} \right)\ ,
  \end{aligned}
  \end{equation}
  schematically shown in Fig.~\ref{fig:2}(a), where $\ket{\pm\alpha}$ designates a coherent state with amplitude $\pm\alpha = \pm \Omega_\text{SB} t /2$.
  The probability $P_{\ket{0}}$ of measuring the initial qubit state $\ket{0}$ then decays to $1/2$ as $\alpha$ increases, following
  \begin{equation} \label{eq:Gaussian}
      \begin{aligned}
      P_{\ket{0}} &= \frac{1}{4}\left(\bra{\alpha} + \bra{-\alpha}\right)\mathds{1}\left(\ket{\alpha} + \ket{-\alpha}\right)\\
      &=\frac{1}{2}\left(1+e^{-2|\alpha|^2}\right)\ ,
  \end{aligned}
  \end{equation}
  providing a measurement of the SDD (and later its suppression).
  For simplicity, we describe the state $\ket{\psi}$ as pure, but in addition to initial thermal population, coherence is degraded by the motional mode heating rate of $\approx$370 quanta/s, comparable to the sideband interaction strength $\Omega_\text{SB} / 2\pi=380$ Hz.
  Loss of coherence increases the rate at which $P_{\ket{0}}$ decays, but the physical intuition presented here remains valid, and the measurement still provides a measurement of $\alpha$ once the independently measured heating rate and initial motional mode occupation are considered (see Sec. S3).
  The qubit state is read out by transferring $\ket{0}$ to the ``dark'' 3$^{2}$D$_\text{5/2}$ manifold and measuring the probability of the ion fluorescing~\cite{myerson2008}.
  A measurement of the decay of $P_{\ket{0}}$ to 0.5, as a consequence of the SDF, is shown in Fig.~\ref{fig:2}(b).

  Driving the system at both DD and sideband frequencies, and varying their relative phases, allows us to suppress the SDD.
  To demonstrate this, the qubit state is measured after a pulse duration which, at most, will reduce $P_{\ket{0}}$ to 0.6 when the DD drive commutes with the SDF.
  To ensure that Rabi oscillations induced by DD driving do not have an impact on the final qubit state (independently of the SDF), we switch the phase of the DD drive in a Walsh-3 pattern \cite{Hayes2012Walsh} (see Sec. S4).
  The resulting data, shown in Fig.~\ref{fig:2}(c), clearly shows that the SDD can be either undisturbed or suppressed by selecting DD phases 0 or $\pi/2$ respectively.
  For this demonstration, we fix the DD drive power injected into the trap such that the DD drive amplitude $\Omega_\text{DD} = 60~\Omega_\text{SB}$, which fulfills the requirement for suppression $\Omega_\text{DD}\gg\Omega_\text{SB}$ (see Sec. S4).
  Note that the microwave power of the dynamical decoupling tone, required to obtain $\Omega_\text{DD} = 60~\Omega_\text{SB}$, is two orders of magnitude smaller than the power injected at sideband frequencies due to the small effective Lamb-Dicke parameter $\eta = 1.25 \times 10^{-3}$ (see definition in Sec. S1).

  \begin{figure}[h]
      \centering
      \includegraphics[width=0.45\textwidth]{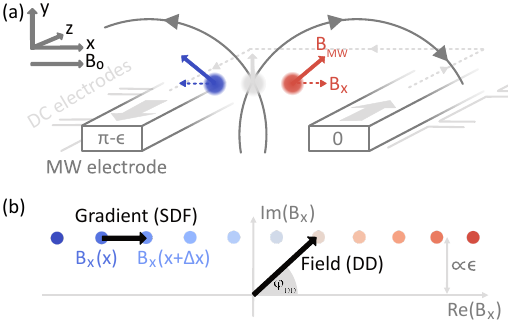}
      \caption{
      \textbf{Surface trap design giving a spatially-varying phase.}
      \textbf{(a)} Schematic of our surface trap.
      Two microwave electrodes produce interfering microwave fields.
      These fields combine to produce $\text{B}_{\text{MW}}$ (solid arrow).
      The component $\text{B}_{\text{x}}$ of this field (dashed arrow) -- in the direction of the quantisation axis, defined by a static field $\text{B}_{0}$ -- drives qubit transitions.
      The two microwave electrodes are connected 750 $\upmu$m away from the ion (grey dashed line) and their currents are approximately $\pi$ radians out-of-phase (with a small phase offset $\epsilon$).
      \textbf{(b)} $\text{B}_{\text{x}}$ phasor in the complex plane (dots) where the imaginary-axis has been magnified $\times 20$.
      Neighboring dots correspond to ion positions $2~\upmu \text{m}$ apart in the x-direction.
      As the ion is moved along the x-axis, the phase of the microwave field, dictating the phase $\phi_{\text{DD}}$, changes by $\approx\pi$ radians.
      The phase of the field gradient, however, remains constant and thus so does that of the state-dependent force.
      The change in phase of the field is used to obtain suppression of the state-dependent displacement away from the symmetry point of the microwave field.
      }
      \label{fig:3}
  \end{figure}

  \begin{figure}[h]
      \centering
      \includegraphics[width=0.45\textwidth]{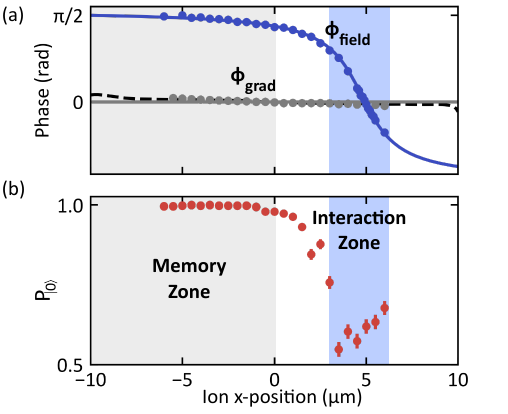}
      \caption{
      \textbf{Spatially-selective ion-motion interaction.}
      \textbf{(a)} Measured phase of the microwave field (blue dots) and its gradient (grey dots), with fits (solid lines) to a Biot-Savart model of the trap (see Sec. S2).
      Here $x=0$ corresponds to the null of the RF trapping field.
      The two shaded regions correspond to the microwave field and its gradient being $\pi/2$ radians out-of-phase (grey) or in-phase (blue).
      The gradient of the microwave field features a slight change in its phase caused by a tilt in the direction of the ion's harmonic motion, consistent with simulated values (black dashed line).
      \textbf{(b)} After preparing state $\ket{0}$ at position $x=0$, the ion is displaced to position $x$, subjected to SB and DD driving, and returned to position $x=0$ where we measure the probability $P_{\ket{0}}$ of finding the qubit in its initial state.
      The phase of the DD drive injected into the trap is kept constant, but the MW electrode geometry offsets the DD phase at the ion position as shown in panel (a).
      This leads to a memory zone (grey) where the state-dependent displacement is suppressed and an interaction zone (blue), where it isn't.
      Here, the interaction zone is defined as where $P_{\ket{0}}$ is below 0.75.
      All error bars indicate 68$\%$ confidence intervals.
      }
      \label{fig:4}
  \end{figure}

  \begin{figure}[h]
      \centering
      \includegraphics[width=0.45\textwidth]{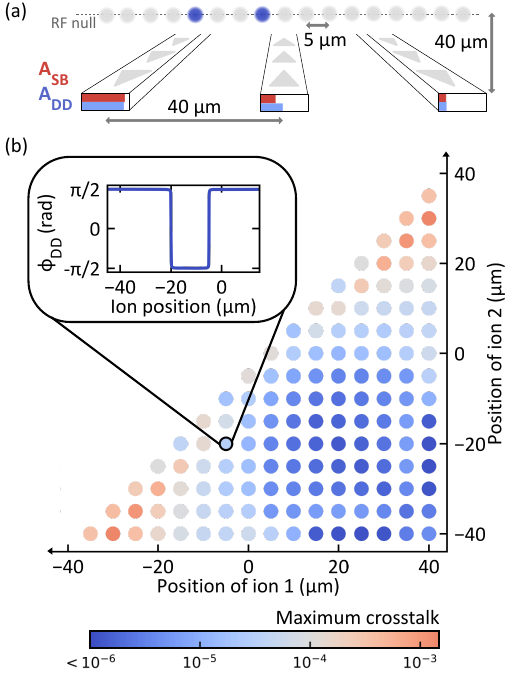}
      \caption{
      \textbf{Simulated surface trap with three microwave electrodes.}
      \textbf{(a)} Schematic of the proposed trap design using three microwave electrodes.
      The ion-chain lies perpendicular to the microwave electrodes.
      A pair of ions (dark blue) have been selected to demonstrate this scheme.
      Bar plots on each electrode show the required sideband (red) and dynamical decoupling (blue) amplitudes to address this ion pair.
      \textbf{(b)} Maximum crosstalk -- due to imperfect suppression of the effect of the state-dependent force (SDF) at the non-addressed ions -- when driving an SDF on both ions in a pair simultaneously.
      The DD phase (corresponding to the microwave field phase) is shown for the ion pair selected in (a).
      }
      \label{fig:5}
  \end{figure}

  To use this technique to address individual ions, the phase difference between the DD driving and SDF at different ion positions must vary -- in our system this arises from the microwave electrode geometry, schematically shown in Fig.~\ref{fig:3}(a).
  The ion is driven by near-field radiation generated by MW currents flowing in a U-shaped electrode (see Ref.~\cite{weber2022cryogenic} for more details).
  Currents propagate in opposite directions on each side of the ion (approximately $\pi$ radians out-of-phase), leading to destructive interference of the magnetic field component $B_x$ aligned with the quantisation axis $B_{0}$ (see Fig.~\ref{fig:3}(a) for coordinate system), which is the component of the field that couples to our ``$\pi$-polarised'' qubit transition.
  The interference leads to a change in the phase of $B_x$ as a function of $x$, since the field switches from being dominated by the field of one electrode to the other.
  This is illustrated by plotting $B_x$ in the complex plane, see Fig.~\ref{fig:3}(b).
  The phase of the \textit{gradient} of the B-field however, which determines the phase of the SDF (see Sec. S1), stays constant as the ion is moved across the trap.
  The phase of the DD drive therefore changes with position, undergoing a phase shift of $\approx \pi$ radians, whereas the phase of the SDF remains constant.
  As a result, the DD and SDF can be in-phase at the x-position of the field minimum whilst being $\approx\pm\pi/2$ out of phase away from it.
  An experimental verification of this is shown in Fig.~\ref{fig:4}(a).
  The change in DD drive phase is measured by preparing (and measuring) the $\ket{+}$ state at $x=0$ before (and after) displacing the ion to position $x$, where it is subject to a MW pulse with varying phase $\phi_p$.
  When averaged over multiple random pulse amplitudes, sweeping the phase $\phi_p$ to maximise the probability of measuring $\ket{+}$ at the end of the sequence constitutes a measurement of the field's position-dependent phase shift (see Sec. S2).
  The phase of the SDF is measured relative to the DD by maximising $P_{\ket{0}}$ in a DD phase scan, as described in Fig.~\ref{fig:2}(c), for different ion positions.
  Ion displacement is achieved by varying the voltage delivered to the trap DC electrodes following an analytical model of the trapping field (see Ref.~\cite{house2008}).
  Notably, we measure a slight change in the SDF phase -- rather than the desired constant phase -- which arises from a change in the direction of the ion's harmonic motion, which samples the microwave gradient in the $y$ direction.
  This results from the limited control offered by our DC trap electrode geometry when displacing the ion from the trap RF null (see Sec. S3).
  The spatial variation in MW phase will ``enable'' the SDD for an ion displaced by $x\approx5~\upmu$m from the trap RF null but the SDD will be increasingly suppressed by the DD drive when the ion is displaced out of this ``interaction zone'', creating a much broader ``memory zone''.
  Experimental verification of this is carried out as in Fig.\ref{fig:2}(c), but where the phase of the DD tone injected in the trap is kept constant, and the ion position is varied instead.
  To disentangle the change in relative phase of the DD from changes in motional mode properties, we vary the pulse duration and sideband frequency with position.
  (We refer here to changes in motional mode frequency, thermal occupation or heating rate, as well as the change in direction of motion, which are measured and presented in Sec. S3 but which would not be present in applications of this technique where ions are not displaced, see Fig.~\ref{fig:5}.)
  The pulse duration is chosen such that, starting in state $\ket{0}$ and in the absence of DD driving, the SDF causes a decay of $P_{\ket{0}}$ to 0.51.
  The measurements, shown in Fig.~\ref{fig:4}(b), reveal the variation in the effect of the state-dependent force as an ion is moved across the surface trap, and the resulting ``interaction'' and ``memory'' zones.

  This technique, extended to two interaction zones, could be used to drive two-qubit gates between arbitrary pairs of ions arranged in a chain, without resorting to ion shuttling.
  We propose a different trap design which would facilitate this, shown in Fig.~\ref{fig:5}(a), where the chain of ions is now perpendicular to the microwave electrodes.
  This could be accomplished by using a multi-layer surface trap \cite{hahn2019multilayer}, possibly where buried microwave electrodes emerge under the ions~\cite{AudeCraik2014}.
  Here the additional interaction zone is enabled by a third microwave electrode creating a second DD phase-flip (at a different position to the first).
  The three electrodes are assumed to be fed by independent microwave currents, where the central MW current is close to $\pi$ radians out of phase with the other two.
  For our simulations, we set a phase difference $(\pi - \epsilon)$ radians, where $\epsilon = 3\times 10^{-5}$ (based on a typical figure for the demonstrated accuracy in MW phase control~\cite{Warring2013Phase}).
  The phase offset $\epsilon$ sets the ``steepness'' of the phase gradient, and therefore the width of the interaction zone.
  The positions of the two phase-flips, and corresponding interaction zones, can be set through the amplitudes of the DD currents through each electrode.
  The SDF strength at both zones can also be set independently by varying the SB currents through each electrode.
  We simulate the expected crosstalk when performing two-qubit gates between all pairs in a chain of 17 ions, with a uniform ion spacing of $5~\upmu\text{m}$ and an ion height of $40~\upmu\text{m}$.
  We assume that the leading source of crosstalk would be imperfect control over the spatial variation in DD phase, i.e. $\phi_\text{DD}-\phi_\text{SB} \ne \pm \pi/2$ and neglect other sources of crosstalk (see Sec. S5).
  Crosstalk here measures the impurity of the partial trace of non-addressed ions after a gate.
  In Fig.~\ref{fig:5}(b), we consider the worst affected non-addressed ion for a given pair of addressed ions and so, since we are not showing the average crosstalk across the chain, we refer to this as ``maximum crosstalk''.
  Alternatively, if we consider the mean crosstalk across the chain, we find a crosstalk error of $1.04(9)\times 10^{-5}$ (averaging over all addressed ion pairs).
  This shows that all-to-all connectivity is possible with crosstalk error rates (due to imperfect SDD suppression at the non-addressed ions) far below typical error correction thresholds.
  Errors are lowest between the electrodes -- the natural location for creating field interference -- but increase as ion pairs are moved further out.
  Adding more microwave electrodes or bringing the phase difference between currents closer to $\pi$ radians reduces these errors and would enable even larger registers of ions.
  For this design, we have considered the chain of ions to be perpendicular to the MW electrodes, for an implementation of gates on axial motional modes.
  However, changing the angle between the ion chain and MW electrodes would also enable the use of radial motional modes.
  As the ion chain gets closer to being parallel with the MW electrodes, the radial MW gradient increases, and narrower interaction zones are required.
  Alternatively, one could use separate electrodes for dynamical decoupling (perpendicular to the ion chain) and sideband driving (parallel to the ion chain, maximising the radial MW gradient).
  Neither electrode direction should however be perpendicular to the quantisation axis.
  In practice, the combination of low ion height and long chains is likely to exacerbate issues of anomalous motional heating~\cite{Brownnutt2015} and limited ion lifetime~\cite{Vittorini2013}.
  To mitigate these, cryogenic operation will probably be required~\cite{weber2022cryogenic}.
  To further limit the impact of heating, the use of out-of-phase motional modes (rather than center-of-mass-modes) should be favoured~\cite{Brownnutt2015}.
  If these issues are too severe, mode frequencies can be raised (reducing inter-ion spacing for axial modes), or the ion height can be increased~\cite{Brownnutt2015}.
  However, such solutions would require finer microwave phase control (more specifically control over $\epsilon$) to maintain low cross-talk, and would reduce gate speeds.
  Other issues with long chains are shared with previous implementations, for example: motional frequency crowding, requiring more complex amplitude and phase shaping~\cite{ShiLiang2006, choi2014}, and the reduction in gate speed resulting from the increase in effective mode mass.
  Gate fidelities will also be affected by drifts in the ion's position~\cite{Kranzl2022}, and we expect that active stabilisation will be required to combat this issue, as discussed in Sec. S5.
  These considerations will ultimately determine the optimal number of ions to use in a chain, and will be influenced by progress in reducing anomalous heating, stabilising the ion's position and controlling the relative microwave phase between on-chip electrodes.
  \begin{figure} [t]
      \centering
      \includegraphics[width=0.45\textwidth]{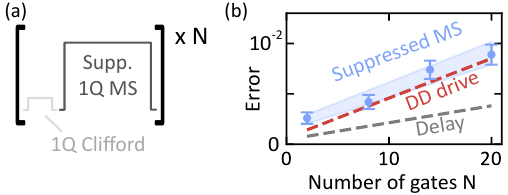}
      \caption{
          \textbf{Single-qubit randomised benchmarking of the suppression error during a gate.}
          \textbf{(a)} The residual error after suppressing the SDD during a gate is experimentally estimated by embedding M\o{}lmer-S\o{}rensen (MS) gate pulses, applied to a single ion and suppressed through a DD drive ($\phi_\text{car}=\pi/2$), in a single-qubit randomised benchmarking (RB) sequence.
          \textbf{(b)} An error of $3.7(4)\times 10^{-4}$ is demonstrated (blue), indistinguishable from the error induced by DD driving alone (red dashed line) indicating a complete suppression of the effect of the ion-motion interaction.
      The DD driving error is partially accounted for by laser leakage and qubit decoherence, measured by replacing the MS pulse by a delay of equal duration (grey dashed line) (see Sec. S6).
      Error bars indicate 68$\%$ confidence intervals of the measurements and the shaded region indicates a 68$\%$ confidence interval of the fit.
      }
      \label{fig:6}
  \end{figure}

  We measure to what extent the theoretical crosstalk would be dominated by other effects in our system by driving a ``suppressed'' M\o{}lmer-S\o{}rensen (MS) interaction in a randomised-benchmarking type measurement \cite{Knill2008RBM} (see Sec. S6).
  We detune the sidebands by $2\pi\times 770$ Hz in a 1.30 ms pulse to emulate -- on a single ion radial mode -- a 1-loop MS gate for a Rabi frequency of $\Omega_\text{SB}=2\pi\times380$ Hz, and apply a DD drive aiming to suppress the effect of this interaction.
  Due to the large heating rate of this mode, an un-suppressed motional interaction is expected to completely decohere the qubit.
  The ``suppressed MS gate'' is applied after every Clifford gate of a single-qubit randomised benchmarking sequence.
  We first run a test measurement with $\phi_\text{DD}=0$, to verify the effectiveness of the RB sequences in revealing motional interaction.
  As expected, we find that the error signal rises to its maximum value (0.5), indicating that the MS-pulses lead to a complete loss of coherence as a result of motional interaction.
  We then change the DD phase to $\phi_\text{DD}=\pi/2$ to suppress the effect of the MS interaction, and obtain an error per ``suppressed MS gate'' of $3.7(4)\times 10^{-4}$, shown in Fig.~\ref{fig:6}.
  DD driving alone produces the same error, $4.0(6)\times 10^{-4}$, suggesting that residual qubit-motion entanglement is negligible compared to this measured error, which is probably of technical origin.
  This error is still far below typical error-correcting thresholds, suggesting strong potential for scaling to multi-ion registers.
  Finally, we measure the error arising if the MS-pulse is replaced by an equally long time delay, revealing that about half of the error, $1.7(2)\times 10^{-4}$, is not related to the microwave driving at all.
  In conclusion, we have demonstrated a method to perform individually-addressed gate interactions with a spatial-resolution finer than typical inter-ion-spacings.
  To do so, we make use of (1) state-dependent displacement (SDD) suppression using dynamical decoupling (DD), and (2) a microwave electrode geometry which generates interference, bringing the DD drive in- and out-of-phase with the state-dependent force (SDF).
  Building upon these concepts, we have proposed a surface trap design to create and move two interaction zones over a chain of ions.
  We predict that this will enable all-to-all connectivity between ions with crosstalk levels far below error-correction thresholds.
  The crosstalk error of $3.7(4)\times 10^{-4}$, emulated in a single-ion benchmarking experiment, supports this prediction.

  This technique could make microwave-driven logic a more practical approach to constructing a large-scale universal quantum processor by reducing its reliance on shuttling ions.
  Once two-qubit gates implemented on a chain of ions become limited by the length of the chain, for example due to multi-mode effects~\cite{Cetina2022,Landsman2019}, one could then rely on shuttling between different logical zones~\cite{Wineland2002QCCD} or optical networking~\cite{Monroe2014Networking}.
  Additionally, this method is compatible with pulse amplitude and phase shaping techniques commonly employed to mitigate errors associated with the many motional modes present in long ion chains \cite{ShiLiang2006, choi2014}, so long as the amplitude ratio and phase difference between the DD and SDF drives are kept constant.
  The proposed microwave electrode geometry can also generate a microwave amplitude gradient enabling addressed single-qubit gates with very low ($< 10^{-4}$) errors~\cite{leu2023}, which, combined with the method for addressed two-qubit gates presented in this report, forms a universal, addressed, gate set.
  Finally, the technique could find applications in a "quantum CCD" architecture~\cite{Wineland2002QCCD}, where entire registers of ions could be dynamically decoupled from gate interactions to mitigate crosstalk problems.

  \vspace{5mm}
  \textbf{Acknowledgments}
  This work was supported by the U.S. Army Research Office (ref. W911NF-18-1-0340) and the U.K. EPSRC Quantum Computing and Simulation Hub.
  M.C.S. acknowledges support from Balliol College.
  A.D.L. acknowledges support from Oxford Ionics Ltd.
  M.F.G. acknowledges support from the Netherlands Organization for Scientific Research (NWO) through a Rubicon Grant.

  \textbf{Author Contributions}
  M.C.S., A.D.L. and M.F.G. contributed equally to this work.
  A.D.L. carried out simulations of the scheme for this experiment and larger ion registers.
  M.C.S. and M.F.G. acquired and analysed the data.
  M.C.S. and M.F.G. wrote the manuscript with contributions from all authors.
  M.C.S., A.D.L. and M.F.G. upgraded and maintained the experiment.
  D.M.L. and M.F.G. supervised the project.

\bibliography{main}

\begin{thebibliography}{60}%
\makeatletter
\providecommand \@ifxundefined [1]{%
 \@ifx{#1\undefined}
}%
\providecommand \@ifnum [1]{%
 \ifnum #1\expandafter \@firstoftwo
 \else \expandafter \@secondoftwo
 \fi
}%
\providecommand \@ifx [1]{%
 \ifx #1\expandafter \@firstoftwo
 \else \expandafter \@secondoftwo
 \fi
}%
\providecommand \natexlab [1]{#1}%
\providecommand \enquote  [1]{``#1''}%
\providecommand \bibnamefont  [1]{#1}%
\providecommand \bibfnamefont [1]{#1}%
\providecommand \citenamefont [1]{#1}%
\providecommand \href@noop [0]{\@secondoftwo}%
\providecommand \href [0]{\begingroup \@sanitize@url \@href}%
\providecommand \@href[1]{\@@startlink{#1}\@@href}%
\providecommand \@@href[1]{\endgroup#1\@@endlink}%
\providecommand \@sanitize@url [0]{\catcode `\\12\catcode `\$12\catcode
  `\&12\catcode `\#12\catcode `\^12\catcode `\_12\catcode `\%12\relax}%
\providecommand \@@startlink[1]{}%
\providecommand \@@endlink[0]{}%
\providecommand \url  [0]{\begingroup\@sanitize@url \@url }%
\providecommand \@url [1]{\endgroup\@href {#1}{\urlprefix }}%
\providecommand \urlprefix  [0]{URL }%
\providecommand \Eprint [0]{\href }%
\providecommand \doibase [0]{https://doi.org/}%
\providecommand \selectlanguage [0]{\@gobble}%
\providecommand \bibinfo  [0]{\@secondoftwo}%
\providecommand \bibfield  [0]{\@secondoftwo}%
\providecommand \translation [1]{[#1]}%
\providecommand \BibitemOpen [0]{}%
\providecommand \bibitemStop [0]{}%
\providecommand \bibitemNoStop [0]{.\EOS\space}%
\providecommand \EOS [0]{\spacefactor3000\relax}%
\providecommand \BibitemShut  [1]{\csname bibitem#1\endcsname}%
\let\auto@bib@innerbib\@empty
\bibitem [{\citenamefont {Viola}\ and\ \citenamefont
  {Lloyd}(1998)}]{viola1998}%
  \BibitemOpen
  \bibfield  {author} {\bibinfo {author} {\bibfnamefont {L.}~\bibnamefont
  {Viola}}\ and\ \bibinfo {author} {\bibfnamefont {S.}~\bibnamefont {Lloyd}},\
  }\bibfield  {title} {\bibinfo {title} {Dynamical suppression of decoherence
  in two-state quantum systems},\ }\href
  {https://doi.org/10.1103/PhysRevA.58.2733} {\bibfield  {journal} {\bibinfo
  {journal} {Phys. Rev. A}\ }\textbf {\bibinfo {volume} {58}},\ \bibinfo
  {pages} {2733} (\bibinfo {year} {1998})}\BibitemShut {NoStop}%
\bibitem [{\citenamefont {Viola}\ \emph {et~al.}(1999)\citenamefont {Viola},
  \citenamefont {Knill},\ and\ \citenamefont {Lloyd}}]{viola1999}%
  \BibitemOpen
  \bibfield  {author} {\bibinfo {author} {\bibfnamefont {L.}~\bibnamefont
  {Viola}}, \bibinfo {author} {\bibfnamefont {E.}~\bibnamefont {Knill}},\ and\
  \bibinfo {author} {\bibfnamefont {S.}~\bibnamefont {Lloyd}},\ }\bibfield
  {title} {\bibinfo {title} {Dynamical decoupling of open quantum systems},\
  }\href {https://doi.org/10.1103/PhysRevLett.82.2417} {\bibfield  {journal}
  {\bibinfo  {journal} {Phys. Rev. Lett.}\ }\textbf {\bibinfo {volume} {82}},\
  \bibinfo {pages} {2417} (\bibinfo {year} {1999})}\BibitemShut {NoStop}%
\bibitem [{\citenamefont {Hahn}(1950)}]{hahn1950}%
  \BibitemOpen
  \bibfield  {author} {\bibinfo {author} {\bibfnamefont {E.~L.}\ \bibnamefont
  {Hahn}},\ }\bibfield  {title} {\bibinfo {title} {Spin echoes},\ }\href
  {https://doi.org/10.1103/PhysRev.80.580} {\bibfield  {journal} {\bibinfo
  {journal} {Phys. Rev.}\ }\textbf {\bibinfo {volume} {80}},\ \bibinfo {pages}
  {580} (\bibinfo {year} {1950})}\BibitemShut {NoStop}%
\bibitem [{\citenamefont {Carr}\ and\ \citenamefont
  {Purcell}(1954)}]{carr1954}%
  \BibitemOpen
  \bibfield  {author} {\bibinfo {author} {\bibfnamefont {H.~Y.}\ \bibnamefont
  {Carr}}\ and\ \bibinfo {author} {\bibfnamefont {E.~M.}\ \bibnamefont
  {Purcell}},\ }\bibfield  {title} {\bibinfo {title} {Effects of diffusion on
  free precession in nuclear magnetic resonance experiments},\ }\href
  {https://doi.org/10.1103/PhysRev.94.630} {\bibfield  {journal} {\bibinfo
  {journal} {Phys. Rev.}\ }\textbf {\bibinfo {volume} {94}},\ \bibinfo {pages}
  {630} (\bibinfo {year} {1954})}\BibitemShut {NoStop}%
\bibitem [{\citenamefont {Meiboom}\ and\ \citenamefont
  {Gill}(1958)}]{meiboom1958}%
  \BibitemOpen
  \bibfield  {author} {\bibinfo {author} {\bibfnamefont {S.}~\bibnamefont
  {Meiboom}}\ and\ \bibinfo {author} {\bibfnamefont {D.}~\bibnamefont {Gill}},\
  }\bibfield  {title} {\bibinfo {title} {Modified spin‐echo method for
  measuring nuclear relaxation times},\ }\href
  {https://doi.org/https://doi.org/10.1063/1.1716296} {\bibfield  {journal}
  {\bibinfo  {journal} {Rev. Sci. Instrum.}\ }\textbf {\bibinfo {volume}
  {29}},\ \bibinfo {pages} {688} (\bibinfo {year} {1958})}\BibitemShut
  {NoStop}%
\bibitem [{\citenamefont {Suter}\ and\ \citenamefont
  {\'Alvarez}(2016)}]{suter2016}%
  \BibitemOpen
  \bibfield  {author} {\bibinfo {author} {\bibfnamefont {D.}~\bibnamefont
  {Suter}}\ and\ \bibinfo {author} {\bibfnamefont {G.~A.}\ \bibnamefont
  {\'Alvarez}},\ }\bibfield  {title} {\bibinfo {title} {Colloquium: Protecting
  quantum information against environmental noise},\ }\href
  {https://doi.org/10.1103/RevModPhys.88.041001} {\bibfield  {journal}
  {\bibinfo  {journal} {Rev. Mod. Phys.}\ }\textbf {\bibinfo {volume} {88}},\
  \bibinfo {pages} {041001} (\bibinfo {year} {2016})}\BibitemShut {NoStop}%
\bibitem [{\citenamefont {Ezzell}\ \emph {et~al.}(2023)\citenamefont {Ezzell},
  \citenamefont {Pokharel}, \citenamefont {Tewala}, \citenamefont {Quiroz},\
  and\ \citenamefont {Lidar}}]{ezzell2023dynamical}%
  \BibitemOpen
  \bibfield  {author} {\bibinfo {author} {\bibfnamefont {N.}~\bibnamefont
  {Ezzell}}, \bibinfo {author} {\bibfnamefont {B.}~\bibnamefont {Pokharel}},
  \bibinfo {author} {\bibfnamefont {L.}~\bibnamefont {Tewala}}, \bibinfo
  {author} {\bibfnamefont {G.}~\bibnamefont {Quiroz}},\ and\ \bibinfo {author}
  {\bibfnamefont {D.~A.}\ \bibnamefont {Lidar}},\ }\bibfield  {title} {\bibinfo
  {title} {Dynamical decoupling for superconducting qubits: A performance
  survey},\ }\href {https://doi.org/10.1103/PhysRevApplied.20.064027}
  {\bibfield  {journal} {\bibinfo  {journal} {Phys. Rev. Appl.}\ }\textbf
  {\bibinfo {volume} {20}},\ \bibinfo {pages} {064027} (\bibinfo {year}
  {2023})}\BibitemShut {NoStop}%
\bibitem [{\citenamefont {Biercuk}\ \emph {et~al.}(2009)\citenamefont
  {Biercuk}, \citenamefont {Uys}, \citenamefont {VanDevender}, \citenamefont
  {Shiga}, \citenamefont {Itano},\ and\ \citenamefont
  {Bollinger}}]{Biercuk2009}%
  \BibitemOpen
  \bibfield  {author} {\bibinfo {author} {\bibfnamefont {M.~J.}\ \bibnamefont
  {Biercuk}}, \bibinfo {author} {\bibfnamefont {H.}~\bibnamefont {Uys}},
  \bibinfo {author} {\bibfnamefont {A.~P.}\ \bibnamefont {VanDevender}},
  \bibinfo {author} {\bibfnamefont {N.}~\bibnamefont {Shiga}}, \bibinfo
  {author} {\bibfnamefont {W.~M.}\ \bibnamefont {Itano}},\ and\ \bibinfo
  {author} {\bibfnamefont {J.~J.}\ \bibnamefont {Bollinger}},\ }\bibfield
  {title} {\bibinfo {title} {Optimized dynamical decoupling in a model quantum
  memory},\ }\href {https://doi.org/10.1038/nature07951} {\bibfield  {journal}
  {\bibinfo  {journal} {Nature}\ }\textbf {\bibinfo {volume} {458}},\ \bibinfo
  {pages} {996} (\bibinfo {year} {2009})}\BibitemShut {NoStop}%
\bibitem [{\citenamefont {Damodarakurup}\ \emph {et~al.}(2009)\citenamefont
  {Damodarakurup}, \citenamefont {Lucamarini}, \citenamefont {Di~Giuseppe},
  \citenamefont {Vitali},\ and\ \citenamefont {Tombesi}}]{Damodarakurup2009}%
  \BibitemOpen
  \bibfield  {author} {\bibinfo {author} {\bibfnamefont {S.}~\bibnamefont
  {Damodarakurup}}, \bibinfo {author} {\bibfnamefont {M.}~\bibnamefont
  {Lucamarini}}, \bibinfo {author} {\bibfnamefont {G.}~\bibnamefont
  {Di~Giuseppe}}, \bibinfo {author} {\bibfnamefont {D.}~\bibnamefont
  {Vitali}},\ and\ \bibinfo {author} {\bibfnamefont {P.}~\bibnamefont
  {Tombesi}},\ }\bibfield  {title} {\bibinfo {title} {Experimental inhibition
  of decoherence on flying qubits via ``bang-bang'' control},\ }\href
  {https://doi.org/10.1103/PhysRevLett.103.040502} {\bibfield  {journal}
  {\bibinfo  {journal} {Phys. Rev. Lett.}\ }\textbf {\bibinfo {volume} {103}},\
  \bibinfo {pages} {040502} (\bibinfo {year} {2009})}\BibitemShut {NoStop}%
\bibitem [{\citenamefont {Du}\ \emph {et~al.}(2009)\citenamefont {Du},
  \citenamefont {Rong}, \citenamefont {Zhao}, \citenamefont {Wang},
  \citenamefont {Yang},\ and\ \citenamefont {Liu}}]{Jiangfeng2009}%
  \BibitemOpen
  \bibfield  {author} {\bibinfo {author} {\bibfnamefont {J.}~\bibnamefont
  {Du}}, \bibinfo {author} {\bibfnamefont {X.}~\bibnamefont {Rong}}, \bibinfo
  {author} {\bibfnamefont {N.}~\bibnamefont {Zhao}}, \bibinfo {author}
  {\bibfnamefont {Y.}~\bibnamefont {Wang}}, \bibinfo {author} {\bibfnamefont
  {J.}~\bibnamefont {Yang}},\ and\ \bibinfo {author} {\bibfnamefont {R.~B.}\
  \bibnamefont {Liu}},\ }\bibfield  {title} {\bibinfo {title} {Preserving
  electron spin coherence in solids by optimal dynamical decoupling},\ }\href
  {https://doi.org/10.1038/nature08470} {\bibfield  {journal} {\bibinfo
  {journal} {Nature}\ }\textbf {\bibinfo {volume} {461}},\ \bibinfo {pages}
  {1265} (\bibinfo {year} {2009})}\BibitemShut {NoStop}%
\bibitem [{\citenamefont {Jenista}\ \emph {et~al.}(2009)\citenamefont
  {Jenista}, \citenamefont {Stokes}, \citenamefont {Branca},\ and\
  \citenamefont {Warren}}]{jenista2009}%
  \BibitemOpen
  \bibfield  {author} {\bibinfo {author} {\bibfnamefont {E.~R.}\ \bibnamefont
  {Jenista}}, \bibinfo {author} {\bibfnamefont {A.~M.}\ \bibnamefont {Stokes}},
  \bibinfo {author} {\bibfnamefont {R.~T.}\ \bibnamefont {Branca}},\ and\
  \bibinfo {author} {\bibfnamefont {W.~S.}\ \bibnamefont {Warren}},\ }\bibfield
   {title} {\bibinfo {title} {Optimized, unequal pulse spacing in multiple echo
  sequences improves refocusing in magnetic resonance},\ }\href
  {https://doi.org/10.1063/1.3263196} {\bibfield  {journal} {\bibinfo
  {journal} {The Journal of Chemical Physics}\ }\textbf {\bibinfo {volume}
  {131}},\ \bibinfo {pages} {204510} (\bibinfo {year} {2009})}\BibitemShut
  {NoStop}%
\bibitem [{\citenamefont {Wang}\ \emph {et~al.}(2012)\citenamefont {Wang},
  \citenamefont {de~Lange}, \citenamefont {Rist\`e}, \citenamefont {Hanson},\
  and\ \citenamefont {Dobrovitski}}]{wang2012}%
  \BibitemOpen
  \bibfield  {author} {\bibinfo {author} {\bibfnamefont {Z.-H.}\ \bibnamefont
  {Wang}}, \bibinfo {author} {\bibfnamefont {G.}~\bibnamefont {de~Lange}},
  \bibinfo {author} {\bibfnamefont {D.}~\bibnamefont {Rist\`e}}, \bibinfo
  {author} {\bibfnamefont {R.}~\bibnamefont {Hanson}},\ and\ \bibinfo {author}
  {\bibfnamefont {V.~V.}\ \bibnamefont {Dobrovitski}},\ }\bibfield  {title}
  {\bibinfo {title} {Comparison of dynamical decoupling protocols for a
  nitrogen-vacancy center in diamond},\ }\href
  {https://doi.org/10.1103/PhysRevB.85.155204} {\bibfield  {journal} {\bibinfo
  {journal} {Phys. Rev. B}\ }\textbf {\bibinfo {volume} {85}},\ \bibinfo
  {pages} {155204} (\bibinfo {year} {2012})}\BibitemShut {NoStop}%
\bibitem [{\citenamefont {Tan}\ \emph {et~al.}(2013)\citenamefont {Tan},
  \citenamefont {Gaebler}, \citenamefont {Bowler}, \citenamefont {Lin},
  \citenamefont {Jost}, \citenamefont {Leibfried},\ and\ \citenamefont
  {Wineland}}]{tan2013}%
  \BibitemOpen
  \bibfield  {author} {\bibinfo {author} {\bibfnamefont {T.~R.}\ \bibnamefont
  {Tan}}, \bibinfo {author} {\bibfnamefont {J.~P.}\ \bibnamefont {Gaebler}},
  \bibinfo {author} {\bibfnamefont {R.}~\bibnamefont {Bowler}}, \bibinfo
  {author} {\bibfnamefont {Y.}~\bibnamefont {Lin}}, \bibinfo {author}
  {\bibfnamefont {J.~D.}\ \bibnamefont {Jost}}, \bibinfo {author}
  {\bibfnamefont {D.}~\bibnamefont {Leibfried}},\ and\ \bibinfo {author}
  {\bibfnamefont {D.~J.}\ \bibnamefont {Wineland}},\ }\bibfield  {title}
  {\bibinfo {title} {Demonstration of a dressed-state phase gate for trapped
  ions},\ }\href {https://doi.org/10.1103/PhysRevLett.110.263002} {\bibfield
  {journal} {\bibinfo  {journal} {Phys. Rev. Lett.}\ }\textbf {\bibinfo
  {volume} {110}},\ \bibinfo {pages} {263002} (\bibinfo {year}
  {2013})}\BibitemShut {NoStop}%
\bibitem [{\citenamefont {Harty}\ \emph {et~al.}(2016)\citenamefont {Harty},
  \citenamefont {Sepiol}, \citenamefont {Allcock}, \citenamefont {Ballance},
  \citenamefont {Tarlton},\ and\ \citenamefont {Lucas}}]{harty2016}%
  \BibitemOpen
  \bibfield  {author} {\bibinfo {author} {\bibfnamefont {T.~P.}\ \bibnamefont
  {Harty}}, \bibinfo {author} {\bibfnamefont {M.~A.}\ \bibnamefont {Sepiol}},
  \bibinfo {author} {\bibfnamefont {D.~T.~C.}\ \bibnamefont {Allcock}},
  \bibinfo {author} {\bibfnamefont {C.~J.}\ \bibnamefont {Ballance}}, \bibinfo
  {author} {\bibfnamefont {J.~E.}\ \bibnamefont {Tarlton}},\ and\ \bibinfo
  {author} {\bibfnamefont {D.~M.}\ \bibnamefont {Lucas}},\ }\bibfield  {title}
  {\bibinfo {title} {High-fidelity trapped-ion quantum logic using near-field
  microwaves},\ }\href {https://doi.org/10.1103/PhysRevLett.117.140501}
  {\bibfield  {journal} {\bibinfo  {journal} {Phys. Rev. Lett.}\ }\textbf
  {\bibinfo {volume} {117}},\ \bibinfo {pages} {140501} (\bibinfo {year}
  {2016})}\BibitemShut {NoStop}%
\bibitem [{\citenamefont {Manovitz}\ \emph {et~al.}(2017)\citenamefont
  {Manovitz}, \citenamefont {Rotem}, \citenamefont {Shaniv}, \citenamefont
  {Cohen}, \citenamefont {Shapira}, \citenamefont {Akerman}, \citenamefont
  {Retzker},\ and\ \citenamefont {Ozeri}}]{manivitz2017}%
  \BibitemOpen
  \bibfield  {author} {\bibinfo {author} {\bibfnamefont {T.}~\bibnamefont
  {Manovitz}}, \bibinfo {author} {\bibfnamefont {A.}~\bibnamefont {Rotem}},
  \bibinfo {author} {\bibfnamefont {R.}~\bibnamefont {Shaniv}}, \bibinfo
  {author} {\bibfnamefont {I.}~\bibnamefont {Cohen}}, \bibinfo {author}
  {\bibfnamefont {Y.}~\bibnamefont {Shapira}}, \bibinfo {author} {\bibfnamefont
  {N.}~\bibnamefont {Akerman}}, \bibinfo {author} {\bibfnamefont
  {A.}~\bibnamefont {Retzker}},\ and\ \bibinfo {author} {\bibfnamefont
  {R.}~\bibnamefont {Ozeri}},\ }\bibfield  {title} {\bibinfo {title} {Fast
  dynamical decoupling of the {M}\o{}lmer-{S}\o{}rensen entangling gate},\
  }\href {https://doi.org/10.1103/PhysRevLett.119.220505} {\bibfield  {journal}
  {\bibinfo  {journal} {Phys. Rev. Lett.}\ }\textbf {\bibinfo {volume} {119}},\
  \bibinfo {pages} {220505} (\bibinfo {year} {2017})}\BibitemShut {NoStop}%
\bibitem [{\citenamefont {Barthel}\ \emph {et~al.}(2023)\citenamefont
  {Barthel}, \citenamefont {Huber}, \citenamefont {Casanova}, \citenamefont
  {Arrazola}, \citenamefont {Niroomand}, \citenamefont {Sriarunothai},
  \citenamefont {Plenio},\ and\ \citenamefont
  {Wunderlich}}]{barthel2022robust}%
  \BibitemOpen
  \bibfield  {author} {\bibinfo {author} {\bibfnamefont {P.}~\bibnamefont
  {Barthel}}, \bibinfo {author} {\bibfnamefont {P.~H.}\ \bibnamefont {Huber}},
  \bibinfo {author} {\bibfnamefont {J.}~\bibnamefont {Casanova}}, \bibinfo
  {author} {\bibfnamefont {I.}~\bibnamefont {Arrazola}}, \bibinfo {author}
  {\bibfnamefont {D.}~\bibnamefont {Niroomand}}, \bibinfo {author}
  {\bibfnamefont {T.}~\bibnamefont {Sriarunothai}}, \bibinfo {author}
  {\bibfnamefont {M.~B.}\ \bibnamefont {Plenio}},\ and\ \bibinfo {author}
  {\bibfnamefont {C.}~\bibnamefont {Wunderlich}},\ }\bibfield  {title}
  {\bibinfo {title} {Robust two-qubit gates using pulsed dynamical
  decoupling},\ }\href {https://doi.org/10.1088/1367-2630/acd4db} {\bibfield
  {journal} {\bibinfo  {journal} {New Journal of Physics}\ }\textbf {\bibinfo
  {volume} {25}},\ \bibinfo {pages} {063023} (\bibinfo {year}
  {2023})}\BibitemShut {NoStop}%
\bibitem [{\citenamefont {Buterakos}\ \emph {et~al.}(2018)\citenamefont
  {Buterakos}, \citenamefont {Throckmorton},\ and\ \citenamefont
  {Das~Sarma}}]{Buterakos2018}%
  \BibitemOpen
  \bibfield  {author} {\bibinfo {author} {\bibfnamefont {D.}~\bibnamefont
  {Buterakos}}, \bibinfo {author} {\bibfnamefont {R.~E.}\ \bibnamefont
  {Throckmorton}},\ and\ \bibinfo {author} {\bibfnamefont {S.}~\bibnamefont
  {Das~Sarma}},\ }\bibfield  {title} {\bibinfo {title} {Crosstalk error
  correction through dynamical decoupling of single-qubit gates in capacitively
  coupled singlet-triplet semiconductor spin qubits},\ }\href
  {https://doi.org/10.1103/PhysRevB.97.045431} {\bibfield  {journal} {\bibinfo
  {journal} {Phys. Rev. B}\ }\textbf {\bibinfo {volume} {97}},\ \bibinfo
  {pages} {045431} (\bibinfo {year} {2018})}\BibitemShut {NoStop}%
\bibitem [{\citenamefont {Tripathi}\ \emph {et~al.}(2022)\citenamefont
  {Tripathi}, \citenamefont {Chen}, \citenamefont {Khezri}, \citenamefont
  {Yip}, \citenamefont {Levenson-Falk},\ and\ \citenamefont
  {Lidar}}]{Tripathi2022}%
  \BibitemOpen
  \bibfield  {author} {\bibinfo {author} {\bibfnamefont {V.}~\bibnamefont
  {Tripathi}}, \bibinfo {author} {\bibfnamefont {H.}~\bibnamefont {Chen}},
  \bibinfo {author} {\bibfnamefont {M.}~\bibnamefont {Khezri}}, \bibinfo
  {author} {\bibfnamefont {K.-W.}\ \bibnamefont {Yip}}, \bibinfo {author}
  {\bibfnamefont {E.}~\bibnamefont {Levenson-Falk}},\ and\ \bibinfo {author}
  {\bibfnamefont {D.~A.}\ \bibnamefont {Lidar}},\ }\bibfield  {title} {\bibinfo
  {title} {Suppression of crosstalk in superconducting qubits using dynamical
  decoupling},\ }\href {https://doi.org/10.1103/PhysRevApplied.18.024068}
  {\bibfield  {journal} {\bibinfo  {journal} {Phys. Rev. Appl.}\ }\textbf
  {\bibinfo {volume} {18}},\ \bibinfo {pages} {024068} (\bibinfo {year}
  {2022})}\BibitemShut {NoStop}%
\bibitem [{\citenamefont {Ospelkaus}\ \emph {et~al.}(2008)\citenamefont
  {Ospelkaus}, \citenamefont {Langer}, \citenamefont {Amini}, \citenamefont
  {Brown}, \citenamefont {Leibfried},\ and\ \citenamefont
  {Wineland}}]{ospelkaus2008}%
  \BibitemOpen
  \bibfield  {author} {\bibinfo {author} {\bibfnamefont {C.}~\bibnamefont
  {Ospelkaus}}, \bibinfo {author} {\bibfnamefont {C.~E.}\ \bibnamefont
  {Langer}}, \bibinfo {author} {\bibfnamefont {J.~M.}\ \bibnamefont {Amini}},
  \bibinfo {author} {\bibfnamefont {K.~R.}\ \bibnamefont {Brown}}, \bibinfo
  {author} {\bibfnamefont {D.}~\bibnamefont {Leibfried}},\ and\ \bibinfo
  {author} {\bibfnamefont {D.~J.}\ \bibnamefont {Wineland}},\ }\bibfield
  {title} {\bibinfo {title} {Trapped-ion quantum logic gates based on
  oscillating magnetic fields},\ }\href
  {https://doi.org/10.1103/PhysRevLett.101.090502} {\bibfield  {journal}
  {\bibinfo  {journal} {Phys. Rev. Lett.}\ }\textbf {\bibinfo {volume} {101}},\
  \bibinfo {pages} {090502} (\bibinfo {year} {2008})}\BibitemShut {NoStop}%
\bibitem [{\citenamefont {Ospelkaus}\ \emph {et~al.}(2011)\citenamefont
  {Ospelkaus}, \citenamefont {Warring}, \citenamefont {Colombe}, \citenamefont
  {Brown}, \citenamefont {Amini}, \citenamefont {Leibfried},\ and\
  \citenamefont {Wineland}}]{ospelkaus2011}%
  \BibitemOpen
  \bibfield  {author} {\bibinfo {author} {\bibfnamefont {C.}~\bibnamefont
  {Ospelkaus}}, \bibinfo {author} {\bibfnamefont {U.}~\bibnamefont {Warring}},
  \bibinfo {author} {\bibfnamefont {Y.}~\bibnamefont {Colombe}}, \bibinfo
  {author} {\bibfnamefont {K.}~\bibnamefont {Brown}}, \bibinfo {author}
  {\bibfnamefont {J.}~\bibnamefont {Amini}}, \bibinfo {author} {\bibfnamefont
  {D.}~\bibnamefont {Leibfried}},\ and\ \bibinfo {author} {\bibfnamefont
  {D.~J.}\ \bibnamefont {Wineland}},\ }\bibfield  {title} {\bibinfo {title}
  {Microwave quantum logic gates for trapped ions},\ }\href
  {https://doi.org/10.1038/nature10290} {\bibfield  {journal} {\bibinfo
  {journal} {Nature}\ }\textbf {\bibinfo {volume} {476}},\ \bibinfo {pages}
  {181} (\bibinfo {year} {2011})}\BibitemShut {NoStop}%
\bibitem [{\citenamefont {Harty}\ \emph {et~al.}(2014)\citenamefont {Harty},
  \citenamefont {Allcock}, \citenamefont {Ballance}, \citenamefont {Guidoni},
  \citenamefont {Janacek}, \citenamefont {Linke}, \citenamefont {Stacey},\ and\
  \citenamefont {Lucas}}]{harty2014}%
  \BibitemOpen
  \bibfield  {author} {\bibinfo {author} {\bibfnamefont {T.~P.}\ \bibnamefont
  {Harty}}, \bibinfo {author} {\bibfnamefont {D.~T.~C.}\ \bibnamefont
  {Allcock}}, \bibinfo {author} {\bibfnamefont {C.~J.}\ \bibnamefont
  {Ballance}}, \bibinfo {author} {\bibfnamefont {L.}~\bibnamefont {Guidoni}},
  \bibinfo {author} {\bibfnamefont {H.~A.}\ \bibnamefont {Janacek}}, \bibinfo
  {author} {\bibfnamefont {N.~M.}\ \bibnamefont {Linke}}, \bibinfo {author}
  {\bibfnamefont {D.~N.}\ \bibnamefont {Stacey}},\ and\ \bibinfo {author}
  {\bibfnamefont {D.~M.}\ \bibnamefont {Lucas}},\ }\bibfield  {title} {\bibinfo
  {title} {High-fidelity preparation, gates, memory, and readout of a
  trapped-ion quantum bit},\ }\href
  {https://doi.org/10.1103/PhysRevLett.113.220501} {\bibfield  {journal}
  {\bibinfo  {journal} {Phys. Rev. Lett.}\ }\textbf {\bibinfo {volume} {113}},\
  \bibinfo {pages} {220501} (\bibinfo {year} {2014})}\BibitemShut {NoStop}%
\bibitem [{\citenamefont {Zarantonello}\ \emph {et~al.}(2019)\citenamefont
  {Zarantonello}, \citenamefont {Hahn}, \citenamefont {Morgner}, \citenamefont
  {Schulte}, \citenamefont {Bautista-Salvador}, \citenamefont {Werner},
  \citenamefont {Hammerer},\ and\ \citenamefont
  {Ospelkaus}}]{zarantonello2019}%
  \BibitemOpen
  \bibfield  {author} {\bibinfo {author} {\bibfnamefont {G.}~\bibnamefont
  {Zarantonello}}, \bibinfo {author} {\bibfnamefont {H.}~\bibnamefont {Hahn}},
  \bibinfo {author} {\bibfnamefont {J.}~\bibnamefont {Morgner}}, \bibinfo
  {author} {\bibfnamefont {M.}~\bibnamefont {Schulte}}, \bibinfo {author}
  {\bibfnamefont {A.}~\bibnamefont {Bautista-Salvador}}, \bibinfo {author}
  {\bibfnamefont {R.~F.}\ \bibnamefont {Werner}}, \bibinfo {author}
  {\bibfnamefont {K.}~\bibnamefont {Hammerer}},\ and\ \bibinfo {author}
  {\bibfnamefont {C.}~\bibnamefont {Ospelkaus}},\ }\bibfield  {title} {\bibinfo
  {title} {Robust and resource-efficient microwave near-field entangling
  $^{9}${B}e$^{+}$ gate},\ }\href
  {https://doi.org/10.1103/PhysRevLett.123.260503} {\bibfield  {journal}
  {\bibinfo  {journal} {Phys. Rev. Lett.}\ }\textbf {\bibinfo {volume} {123}},\
  \bibinfo {pages} {260503} (\bibinfo {year} {2019})}\BibitemShut {NoStop}%
\bibitem [{\citenamefont {Hahn}\ \emph
  {et~al.}(2019{\natexlab{a}})\citenamefont {Hahn}, \citenamefont
  {Zarantonello}, \citenamefont {Schulte}, \citenamefont {Bautista-Salvador},
  \citenamefont {Hammerer},\ and\ \citenamefont {Ospelkaus}}]{Hahn2019a}%
  \BibitemOpen
  \bibfield  {author} {\bibinfo {author} {\bibfnamefont {H.}~\bibnamefont
  {Hahn}}, \bibinfo {author} {\bibfnamefont {G.}~\bibnamefont {Zarantonello}},
  \bibinfo {author} {\bibfnamefont {M.}~\bibnamefont {Schulte}}, \bibinfo
  {author} {\bibfnamefont {A.}~\bibnamefont {Bautista-Salvador}}, \bibinfo
  {author} {\bibfnamefont {K.}~\bibnamefont {Hammerer}},\ and\ \bibinfo
  {author} {\bibfnamefont {C.}~\bibnamefont {Ospelkaus}},\ }\bibfield  {title}
  {\bibinfo {title} {{Integrated 9Be+ multi-qubit gate device for the ion-trap
  quantum computer}},\ }\href {https://doi.org/10.1038/s41534-019-0184-5}
  {\bibfield  {journal} {\bibinfo  {journal} {npj Quantum Information}\
  }\textbf {\bibinfo {volume} {5}},\ \bibinfo {pages} {2} (\bibinfo {year}
  {2019}{\natexlab{a}})},\ \Eprint {https://arxiv.org/abs/1902.07028}
  {arXiv:1902.07028} \BibitemShut {NoStop}%
\bibitem [{\citenamefont {Weidt}\ \emph {et~al.}(2016)\citenamefont {Weidt},
  \citenamefont {Randall}, \citenamefont {Webster}, \citenamefont {Lake},
  \citenamefont {Webb}, \citenamefont {Cohen}, \citenamefont {Navickas},
  \citenamefont {Lekitsch}, \citenamefont {Retzker},\ and\ \citenamefont
  {Hensinger}}]{Weidt2016}%
  \BibitemOpen
  \bibfield  {author} {\bibinfo {author} {\bibfnamefont {S.}~\bibnamefont
  {Weidt}}, \bibinfo {author} {\bibfnamefont {J.}~\bibnamefont {Randall}},
  \bibinfo {author} {\bibfnamefont {S.~C.}\ \bibnamefont {Webster}}, \bibinfo
  {author} {\bibfnamefont {K.}~\bibnamefont {Lake}}, \bibinfo {author}
  {\bibfnamefont {A.~E.}\ \bibnamefont {Webb}}, \bibinfo {author}
  {\bibfnamefont {I.}~\bibnamefont {Cohen}}, \bibinfo {author} {\bibfnamefont
  {T.}~\bibnamefont {Navickas}}, \bibinfo {author} {\bibfnamefont
  {B.}~\bibnamefont {Lekitsch}}, \bibinfo {author} {\bibfnamefont
  {A.}~\bibnamefont {Retzker}},\ and\ \bibinfo {author} {\bibfnamefont {W.~K.}\
  \bibnamefont {Hensinger}},\ }\bibfield  {title} {\bibinfo {title}
  {{Trapped-Ion Quantum Logic with Global Radiation Fields}},\ }\bibfield
  {journal} {\bibinfo  {journal} {Physical Review Letters}\ }\textbf {\bibinfo
  {volume} {117}},\ \href {https://doi.org/10.1103/PhysRevLett.117.220501}
  {10.1103/PhysRevLett.117.220501} (\bibinfo {year} {2016}),\ \Eprint
  {https://arxiv.org/abs/1603.03384} {arXiv:1603.03384} \BibitemShut {NoStop}%
\bibitem [{\citenamefont {Srinivas}\ \emph {et~al.}(2021)\citenamefont
  {Srinivas}, \citenamefont {Burd}, \citenamefont {Knaack}, \citenamefont
  {Sutherland}, \citenamefont {Kwiatkowski}, \citenamefont {Glancy},
  \citenamefont {Knill}, \citenamefont {Wineland}, \citenamefont {Leibfried},
  \citenamefont {Wilson}, \citenamefont {Allcock},\ and\ \citenamefont
  {Slichter}}]{srinivas2021}%
  \BibitemOpen
  \bibfield  {author} {\bibinfo {author} {\bibfnamefont {R.}~\bibnamefont
  {Srinivas}}, \bibinfo {author} {\bibfnamefont {S.~C.}\ \bibnamefont {Burd}},
  \bibinfo {author} {\bibfnamefont {H.~M.}\ \bibnamefont {Knaack}}, \bibinfo
  {author} {\bibfnamefont {R.~T.}\ \bibnamefont {Sutherland}}, \bibinfo
  {author} {\bibfnamefont {A.}~\bibnamefont {Kwiatkowski}}, \bibinfo {author}
  {\bibfnamefont {S.}~\bibnamefont {Glancy}}, \bibinfo {author} {\bibfnamefont
  {E.}~\bibnamefont {Knill}}, \bibinfo {author} {\bibfnamefont {D.~J.}\
  \bibnamefont {Wineland}}, \bibinfo {author} {\bibfnamefont {D.}~\bibnamefont
  {Leibfried}}, \bibinfo {author} {\bibfnamefont {A.~C.}\ \bibnamefont
  {Wilson}}, \bibinfo {author} {\bibfnamefont {D.~T.~C.}\ \bibnamefont
  {Allcock}},\ and\ \bibinfo {author} {\bibfnamefont {D.~H.}\ \bibnamefont
  {Slichter}},\ }\bibfield  {title} {\bibinfo {title} {High-fidelity laser-free
  universal control of trapped ion qubits},\ }\href
  {https://doi.org/10.1038/s41586-021-03809-4} {\bibfield  {journal} {\bibinfo
  {journal} {Nature}\ }\textbf {\bibinfo {volume} {597}},\ \bibinfo {pages}
  {209–213} (\bibinfo {year} {2021})}\BibitemShut {NoStop}%
\bibitem [{\citenamefont {N\"agerl}\ \emph {et~al.}(1999)\citenamefont
  {N\"agerl}, \citenamefont {Leibfried}, \citenamefont {Rohde}, \citenamefont
  {Thalhammer}, \citenamefont {Eschner}, \citenamefont {Schmidt-Kaler},\ and\
  \citenamefont {Blatt}}]{nagerl1999}%
  \BibitemOpen
  \bibfield  {author} {\bibinfo {author} {\bibfnamefont {H.~C.}\ \bibnamefont
  {N\"agerl}}, \bibinfo {author} {\bibfnamefont {D.}~\bibnamefont {Leibfried}},
  \bibinfo {author} {\bibfnamefont {H.}~\bibnamefont {Rohde}}, \bibinfo
  {author} {\bibfnamefont {G.}~\bibnamefont {Thalhammer}}, \bibinfo {author}
  {\bibfnamefont {J.}~\bibnamefont {Eschner}}, \bibinfo {author} {\bibfnamefont
  {F.}~\bibnamefont {Schmidt-Kaler}},\ and\ \bibinfo {author} {\bibfnamefont
  {R.}~\bibnamefont {Blatt}},\ }\bibfield  {title} {\bibinfo {title} {Laser
  addressing of individual ions in a linear ion trap},\ }\href
  {https://doi.org/10.1103/PhysRevA.60.145} {\bibfield  {journal} {\bibinfo
  {journal} {Phys. Rev. A}\ }\textbf {\bibinfo {volume} {60}},\ \bibinfo
  {pages} {145} (\bibinfo {year} {1999})}\BibitemShut {NoStop}%
\bibitem [{\citenamefont {Piltz}\ \emph {et~al.}(2014)\citenamefont {Piltz},
  \citenamefont {Sriarunothai}, \citenamefont {Var{\'o}n},\ and\ \citenamefont
  {Wunderlich}}]{piltz2014}%
  \BibitemOpen
  \bibfield  {author} {\bibinfo {author} {\bibfnamefont {C.}~\bibnamefont
  {Piltz}}, \bibinfo {author} {\bibfnamefont {T.}~\bibnamefont {Sriarunothai}},
  \bibinfo {author} {\bibfnamefont {A.}~\bibnamefont {Var{\'o}n}},\ and\
  \bibinfo {author} {\bibfnamefont {C.}~\bibnamefont {Wunderlich}},\ }\bibfield
   {title} {\bibinfo {title} {A trapped-ion-based quantum byte with $10^{- 5}$
  next-neighbour cross-talk},\ }\href {https://doi.org/10.1038/ncomms5679}
  {\bibfield  {journal} {\bibinfo  {journal} {Nature communications}\ }\textbf
  {\bibinfo {volume} {5}},\ \bibinfo {pages} {4679} (\bibinfo {year}
  {2014})}\BibitemShut {NoStop}%
\bibitem [{\citenamefont {Warring}\ \emph
  {et~al.}(2013{\natexlab{a}})\citenamefont {Warring}, \citenamefont
  {Ospelkaus}, \citenamefont {Colombe}, \citenamefont {J\"ordens},
  \citenamefont {Leibfried},\ and\ \citenamefont {Wineland}}]{warring2013}%
  \BibitemOpen
  \bibfield  {author} {\bibinfo {author} {\bibfnamefont {U.}~\bibnamefont
  {Warring}}, \bibinfo {author} {\bibfnamefont {C.}~\bibnamefont {Ospelkaus}},
  \bibinfo {author} {\bibfnamefont {Y.}~\bibnamefont {Colombe}}, \bibinfo
  {author} {\bibfnamefont {R.}~\bibnamefont {J\"ordens}}, \bibinfo {author}
  {\bibfnamefont {D.}~\bibnamefont {Leibfried}},\ and\ \bibinfo {author}
  {\bibfnamefont {D.~J.}\ \bibnamefont {Wineland}},\ }\bibfield  {title}
  {\bibinfo {title} {Individual-ion addressing with microwave field
  gradients},\ }\href {https://doi.org/10.1103/PhysRevLett.110.173002}
  {\bibfield  {journal} {\bibinfo  {journal} {Phys. Rev. Lett.}\ }\textbf
  {\bibinfo {volume} {110}},\ \bibinfo {pages} {173002} (\bibinfo {year}
  {2013}{\natexlab{a}})}\BibitemShut {NoStop}%
\bibitem [{\citenamefont {Randall}\ \emph {et~al.}(2015)\citenamefont
  {Randall}, \citenamefont {Weidt}, \citenamefont {Standing}, \citenamefont
  {Lake}, \citenamefont {Webster}, \citenamefont {Murgia}, \citenamefont
  {Navickas}, \citenamefont {Roth},\ and\ \citenamefont
  {Hensinger}}]{randell2015}%
  \BibitemOpen
  \bibfield  {author} {\bibinfo {author} {\bibfnamefont {J.}~\bibnamefont
  {Randall}}, \bibinfo {author} {\bibfnamefont {S.}~\bibnamefont {Weidt}},
  \bibinfo {author} {\bibfnamefont {E.~D.}\ \bibnamefont {Standing}}, \bibinfo
  {author} {\bibfnamefont {K.}~\bibnamefont {Lake}}, \bibinfo {author}
  {\bibfnamefont {S.~C.}\ \bibnamefont {Webster}}, \bibinfo {author}
  {\bibfnamefont {D.~F.}\ \bibnamefont {Murgia}}, \bibinfo {author}
  {\bibfnamefont {T.}~\bibnamefont {Navickas}}, \bibinfo {author}
  {\bibfnamefont {K.}~\bibnamefont {Roth}},\ and\ \bibinfo {author}
  {\bibfnamefont {W.~K.}\ \bibnamefont {Hensinger}},\ }\bibfield  {title}
  {\bibinfo {title} {Efficient preparation and detection of microwave
  dressed-state qubits and qutrits with trapped ions},\ }\href
  {https://link.aps.org/doi/10.1103/PhysRevA.91.012322} {\bibfield  {journal}
  {\bibinfo  {journal} {Phys. Rev. A}\ }\textbf {\bibinfo {volume} {91}},\
  \bibinfo {pages} {012322} (\bibinfo {year} {2015})}\BibitemShut {NoStop}%
\bibitem [{\citenamefont {Aude~Craik}\ \emph {et~al.}(2017)\citenamefont
  {Aude~Craik}, \citenamefont {Linke}, \citenamefont {Sepiol}, \citenamefont
  {Harty}, \citenamefont {Goodwin}, \citenamefont {Ballance}, \citenamefont
  {Stacey}, \citenamefont {Steane}, \citenamefont {Lucas},\ and\ \citenamefont
  {Allcock}}]{craik2017}%
  \BibitemOpen
  \bibfield  {author} {\bibinfo {author} {\bibfnamefont {D.~P.~L.}\
  \bibnamefont {Aude~Craik}}, \bibinfo {author} {\bibfnamefont {N.~M.}\
  \bibnamefont {Linke}}, \bibinfo {author} {\bibfnamefont {M.~A.}\ \bibnamefont
  {Sepiol}}, \bibinfo {author} {\bibfnamefont {T.~P.}\ \bibnamefont {Harty}},
  \bibinfo {author} {\bibfnamefont {J.~F.}\ \bibnamefont {Goodwin}}, \bibinfo
  {author} {\bibfnamefont {C.~J.}\ \bibnamefont {Ballance}}, \bibinfo {author}
  {\bibfnamefont {D.~N.}\ \bibnamefont {Stacey}}, \bibinfo {author}
  {\bibfnamefont {A.~M.}\ \bibnamefont {Steane}}, \bibinfo {author}
  {\bibfnamefont {D.~M.}\ \bibnamefont {Lucas}},\ and\ \bibinfo {author}
  {\bibfnamefont {D.~T.~C.}\ \bibnamefont {Allcock}},\ }\bibfield  {title}
  {\bibinfo {title} {High-fidelity spatial and polarization addressing of
  $^{43}\mathrm{Ca}^{+}$ qubits using near-field microwave control},\ }\href
  {https://doi.org/10.1103/PhysRevA.95.022337} {\bibfield  {journal} {\bibinfo
  {journal} {Phys. Rev. A}\ }\textbf {\bibinfo {volume} {95}},\ \bibinfo
  {pages} {022337} (\bibinfo {year} {2017})}\BibitemShut {NoStop}%
\bibitem [{\citenamefont {Sutherland}\ \emph {et~al.}(2023)\citenamefont
  {Sutherland}, \citenamefont {Srinivas},\ and\ \citenamefont
  {Allcock}}]{sutherland2022}%
  \BibitemOpen
  \bibfield  {author} {\bibinfo {author} {\bibfnamefont {R.}~\bibnamefont
  {Sutherland}}, \bibinfo {author} {\bibfnamefont {R.}~\bibnamefont
  {Srinivas}},\ and\ \bibinfo {author} {\bibfnamefont {D.}~\bibnamefont
  {Allcock}},\ }\bibfield  {title} {\bibinfo {title} {Individual addressing of
  trapped ion qubits with geometric phase gates},\ }\href
  {https://doi.org/10.1103/PhysRevA.107.032604} {\bibfield  {journal} {\bibinfo
   {journal} {Phys. Rev. A}\ }\textbf {\bibinfo {volume} {107}},\ \bibinfo
  {pages} {032604} (\bibinfo {year} {2023})}\BibitemShut {NoStop}%
\bibitem [{\citenamefont {Srinivas}\ \emph {et~al.}(2023)\citenamefont
  {Srinivas}, \citenamefont {Löschnauer}, \citenamefont {Malinowski},
  \citenamefont {Hughes}, \citenamefont {Nourshargh}, \citenamefont
  {Negnevitsky}, \citenamefont {Allcock}, \citenamefont {King}, \citenamefont
  {Matthiesen}, \citenamefont {Harty},\ and\ \citenamefont
  {Ballance}}]{srinivas2022}%
  \BibitemOpen
  \bibfield  {author} {\bibinfo {author} {\bibfnamefont {R.}~\bibnamefont
  {Srinivas}}, \bibinfo {author} {\bibfnamefont {C.~M.}\ \bibnamefont
  {Löschnauer}}, \bibinfo {author} {\bibfnamefont {M.}~\bibnamefont
  {Malinowski}}, \bibinfo {author} {\bibfnamefont {A.~C.}\ \bibnamefont
  {Hughes}}, \bibinfo {author} {\bibfnamefont {R.}~\bibnamefont {Nourshargh}},
  \bibinfo {author} {\bibfnamefont {V.}~\bibnamefont {Negnevitsky}}, \bibinfo
  {author} {\bibfnamefont {D.~T.~C.}\ \bibnamefont {Allcock}}, \bibinfo
  {author} {\bibfnamefont {S.~A.}\ \bibnamefont {King}}, \bibinfo {author}
  {\bibfnamefont {C.}~\bibnamefont {Matthiesen}}, \bibinfo {author}
  {\bibfnamefont {T.~P.}\ \bibnamefont {Harty}},\ and\ \bibinfo {author}
  {\bibfnamefont {C.~J.}\ \bibnamefont {Ballance}},\ }\bibfield  {title}
  {\bibinfo {title} {Coherent control of trapped ion qubits with localized
  electric fields},\ }\href {https://doi.org/10.1103/PhysRevLett.131.020601}
  {\bibfield  {journal} {\bibinfo  {journal} {Phys. Rev. Lett.}\ }\textbf
  {\bibinfo {volume} {131}},\ \bibinfo {pages} {020601} (\bibinfo {year}
  {2023})}\BibitemShut {NoStop}%
\bibitem [{\citenamefont {Leu}\ \emph {et~al.}(2023)\citenamefont {Leu},
  \citenamefont {Gely}, \citenamefont {Weber}, \citenamefont {Smith},
  \citenamefont {Nadlinger},\ and\ \citenamefont {Lucas}}]{leu2023}%
  \BibitemOpen
  \bibfield  {author} {\bibinfo {author} {\bibfnamefont {A.~D.}\ \bibnamefont
  {Leu}}, \bibinfo {author} {\bibfnamefont {M.~F.}\ \bibnamefont {Gely}},
  \bibinfo {author} {\bibfnamefont {M.~A.}\ \bibnamefont {Weber}}, \bibinfo
  {author} {\bibfnamefont {M.~C.}\ \bibnamefont {Smith}}, \bibinfo {author}
  {\bibfnamefont {D.~P.}\ \bibnamefont {Nadlinger}},\ and\ \bibinfo {author}
  {\bibfnamefont {D.~M.}\ \bibnamefont {Lucas}},\ }\bibfield  {title} {\bibinfo
  {title} {Fast, high-fidelity addressed single-qubit gates using efficient
  composite pulse sequences},\ }\href
  {https://doi.org/10.1103/PhysRevLett.131.120601} {\bibfield  {journal}
  {\bibinfo  {journal} {Phys. Rev. Lett.}\ }\textbf {\bibinfo {volume} {131}},\
  \bibinfo {pages} {120601} (\bibinfo {year} {2023})}\BibitemShut {NoStop}%
\bibitem [{\citenamefont {Khromova}\ \emph {et~al.}(2012)\citenamefont
  {Khromova}, \citenamefont {Piltz}, \citenamefont {Scharfenberger},
  \citenamefont {Gloger}, \citenamefont {Johanning}, \citenamefont {Var\'on},\
  and\ \citenamefont {Wunderlich}}]{Khromova2012}%
  \BibitemOpen
  \bibfield  {author} {\bibinfo {author} {\bibfnamefont {A.}~\bibnamefont
  {Khromova}}, \bibinfo {author} {\bibfnamefont {C.}~\bibnamefont {Piltz}},
  \bibinfo {author} {\bibfnamefont {B.}~\bibnamefont {Scharfenberger}},
  \bibinfo {author} {\bibfnamefont {T.~F.}\ \bibnamefont {Gloger}}, \bibinfo
  {author} {\bibfnamefont {M.}~\bibnamefont {Johanning}}, \bibinfo {author}
  {\bibfnamefont {A.~F.}\ \bibnamefont {Var\'on}},\ and\ \bibinfo {author}
  {\bibfnamefont {C.}~\bibnamefont {Wunderlich}},\ }\bibfield  {title}
  {\bibinfo {title} {Designer spin pseudomolecule implemented with trapped ions
  in a magnetic gradient},\ }\href
  {https://doi.org/10.1103/PhysRevLett.108.220502} {\bibfield  {journal}
  {\bibinfo  {journal} {Phys. Rev. Lett.}\ }\textbf {\bibinfo {volume} {108}},\
  \bibinfo {pages} {220502} (\bibinfo {year} {2012})}\BibitemShut {NoStop}%
\bibitem [{\citenamefont {S\o{}rensen}\ and\ \citenamefont
  {M\o{}lmer}(2000)}]{sorenson2000}%
  \BibitemOpen
  \bibfield  {author} {\bibinfo {author} {\bibfnamefont {A.}~\bibnamefont
  {S\o{}rensen}}\ and\ \bibinfo {author} {\bibfnamefont {K.}~\bibnamefont
  {M\o{}lmer}},\ }\bibfield  {title} {\bibinfo {title} {Entanglement and
  quantum computation with ions in thermal motion},\ }\href
  {https://doi.org/10.1103/PhysRevA.62.022311} {\bibfield  {journal} {\bibinfo
  {journal} {Phys. Rev. A}\ }\textbf {\bibinfo {volume} {62}},\ \bibinfo
  {pages} {022311} (\bibinfo {year} {2000})}\BibitemShut {NoStop}%
\bibitem [{\citenamefont {S\o{}rensen}\ and\ \citenamefont
  {M\o{}lmer}(1999)}]{sorenson1999}%
  \BibitemOpen
  \bibfield  {author} {\bibinfo {author} {\bibfnamefont {A.}~\bibnamefont
  {S\o{}rensen}}\ and\ \bibinfo {author} {\bibfnamefont {K.}~\bibnamefont
  {M\o{}lmer}},\ }\bibfield  {title} {\bibinfo {title} {Quantum computation
  with ions in thermal motion},\ }\href
  {https://doi.org/10.1103/PhysRevLett.82.1971} {\bibfield  {journal} {\bibinfo
   {journal} {Phys. Rev. Lett.}\ }\textbf {\bibinfo {volume} {82}},\ \bibinfo
  {pages} {1971} (\bibinfo {year} {1999})}\BibitemShut {NoStop}%
\bibitem [{\citenamefont {Milburn}\ \emph {et~al.}(2000)\citenamefont
  {Milburn}, \citenamefont {Schneider},\ and\ \citenamefont
  {James}}]{milburn2000}%
  \BibitemOpen
  \bibfield  {author} {\bibinfo {author} {\bibfnamefont {G.}~\bibnamefont
  {Milburn}}, \bibinfo {author} {\bibfnamefont {S.}~\bibnamefont {Schneider}},\
  and\ \bibinfo {author} {\bibfnamefont {D.}~\bibnamefont {James}},\ }\bibfield
   {title} {\bibinfo {title} {Ion trap quantum computing with warm ions},\
  }\href
  {https://doi.org/https://doi.org/10.1002/1521-3978(200009)48:9/11<801::AID-PROP801>3.0.CO;2-1}
  {\bibfield  {journal} {\bibinfo  {journal} {Fortschritte der Physik}\
  }\textbf {\bibinfo {volume} {48}},\ \bibinfo {pages} {801} (\bibinfo {year}
  {2000})}\BibitemShut {NoStop}%
\bibitem [{\citenamefont {Weber}\ \emph
  {et~al.}(2024{\natexlab{a}})\citenamefont {Weber}, \citenamefont
  {Löschnauer}, \citenamefont {Wolf}, \citenamefont {Gely}, \citenamefont
  {Hanley}, \citenamefont {Goodwin}, \citenamefont {Ballance}, \citenamefont
  {Harty},\ and\ \citenamefont {Lucas}}]{weber2022cryogenic}%
  \BibitemOpen
  \bibfield  {author} {\bibinfo {author} {\bibfnamefont {M.~A.}\ \bibnamefont
  {Weber}}, \bibinfo {author} {\bibfnamefont {C.}~\bibnamefont {Löschnauer}},
  \bibinfo {author} {\bibfnamefont {J.}~\bibnamefont {Wolf}}, \bibinfo {author}
  {\bibfnamefont {M.~F.}\ \bibnamefont {Gely}}, \bibinfo {author}
  {\bibfnamefont {R.~K.}\ \bibnamefont {Hanley}}, \bibinfo {author}
  {\bibfnamefont {J.~F.}\ \bibnamefont {Goodwin}}, \bibinfo {author}
  {\bibfnamefont {C.~J.}\ \bibnamefont {Ballance}}, \bibinfo {author}
  {\bibfnamefont {T.~P.}\ \bibnamefont {Harty}},\ and\ \bibinfo {author}
  {\bibfnamefont {D.~M.}\ \bibnamefont {Lucas}},\ }\bibfield  {title} {\bibinfo
  {title} {Cryogenic ion trap system for high-fidelity near-field
  microwave-driven quantum logic},\ }\href
  {https://doi.org/10.1088/2058-9565/acfba8} {\bibfield  {journal} {\bibinfo
  {journal} {Quantum Sci. Technol.}\ }\textbf {\bibinfo {volume} {9}},\
  \bibinfo {pages} {015007} (\bibinfo {year} {2024}{\natexlab{a}})}\BibitemShut
  {NoStop}%
\bibitem [{\citenamefont {Monroe}\ \emph {et~al.}(1996)\citenamefont {Monroe},
  \citenamefont {Meekhof}, \citenamefont {King},\ and\ \citenamefont
  {Wineland}}]{Wineland1996CatState}%
  \BibitemOpen
  \bibfield  {author} {\bibinfo {author} {\bibfnamefont {C.}~\bibnamefont
  {Monroe}}, \bibinfo {author} {\bibfnamefont {D.~M.}\ \bibnamefont {Meekhof}},
  \bibinfo {author} {\bibfnamefont {B.~E.}\ \bibnamefont {King}},\ and\
  \bibinfo {author} {\bibfnamefont {D.~J.}\ \bibnamefont {Wineland}},\
  }\bibfield  {title} {\bibinfo {title} {A “{S}chrödinger cat”
  superposition state of an atom},\ }\href
  {https://doi.org/10.1126/science.272.5265.1131} {\bibfield  {journal}
  {\bibinfo  {journal} {Science}\ }\textbf {\bibinfo {volume} {272}},\ \bibinfo
  {pages} {1131} (\bibinfo {year} {1996})}\BibitemShut {NoStop}%
\bibitem [{\citenamefont {Myerson}\ \emph {et~al.}(2008)\citenamefont
  {Myerson}, \citenamefont {Szwer}, \citenamefont {Webster}, \citenamefont
  {Allcock}, \citenamefont {Curtis}, \citenamefont {Imreh}, \citenamefont
  {Sherman}, \citenamefont {Stacey}, \citenamefont {Steane},\ and\
  \citenamefont {Lucas}}]{myerson2008}%
  \BibitemOpen
  \bibfield  {author} {\bibinfo {author} {\bibfnamefont {A.~H.}\ \bibnamefont
  {Myerson}}, \bibinfo {author} {\bibfnamefont {D.~J.}\ \bibnamefont {Szwer}},
  \bibinfo {author} {\bibfnamefont {S.~C.}\ \bibnamefont {Webster}}, \bibinfo
  {author} {\bibfnamefont {D.~T.~C.}\ \bibnamefont {Allcock}}, \bibinfo
  {author} {\bibfnamefont {M.~J.}\ \bibnamefont {Curtis}}, \bibinfo {author}
  {\bibfnamefont {G.}~\bibnamefont {Imreh}}, \bibinfo {author} {\bibfnamefont
  {J.~A.}\ \bibnamefont {Sherman}}, \bibinfo {author} {\bibfnamefont {D.~N.}\
  \bibnamefont {Stacey}}, \bibinfo {author} {\bibfnamefont {A.~M.}\
  \bibnamefont {Steane}},\ and\ \bibinfo {author} {\bibfnamefont {D.~M.}\
  \bibnamefont {Lucas}},\ }\bibfield  {title} {\bibinfo {title} {High-fidelity
  readout of trapped-ion qubits},\ }\href
  {https://doi.org/10.1103/PhysRevLett.100.200502} {\bibfield  {journal}
  {\bibinfo  {journal} {Phys. Rev. Lett.}\ }\textbf {\bibinfo {volume} {100}},\
  \bibinfo {pages} {200502} (\bibinfo {year} {2008})}\BibitemShut {NoStop}%
\bibitem [{\citenamefont {Hayes}\ \emph {et~al.}(2012)\citenamefont {Hayes},
  \citenamefont {Clark}, \citenamefont {Debnath}, \citenamefont {Hucul},
  \citenamefont {Inlek}, \citenamefont {Lee}, \citenamefont {Quraishi},\ and\
  \citenamefont {Monroe}}]{Hayes2012Walsh}%
  \BibitemOpen
  \bibfield  {author} {\bibinfo {author} {\bibfnamefont {D.}~\bibnamefont
  {Hayes}}, \bibinfo {author} {\bibfnamefont {S.~M.}\ \bibnamefont {Clark}},
  \bibinfo {author} {\bibfnamefont {S.}~\bibnamefont {Debnath}}, \bibinfo
  {author} {\bibfnamefont {D.}~\bibnamefont {Hucul}}, \bibinfo {author}
  {\bibfnamefont {I.~V.}\ \bibnamefont {Inlek}}, \bibinfo {author}
  {\bibfnamefont {K.~W.}\ \bibnamefont {Lee}}, \bibinfo {author} {\bibfnamefont
  {Q.}~\bibnamefont {Quraishi}},\ and\ \bibinfo {author} {\bibfnamefont
  {C.}~\bibnamefont {Monroe}},\ }\bibfield  {title} {\bibinfo {title} {Coherent
  error suppression in multiqubit entangling gates},\ }\href
  {https://doi.org/10.1103/PhysRevLett.109.020503} {\bibfield  {journal}
  {\bibinfo  {journal} {Phys. Rev. Lett.}\ }\textbf {\bibinfo {volume} {109}},\
  \bibinfo {pages} {020503} (\bibinfo {year} {2012})}\BibitemShut {NoStop}%
\bibitem [{\citenamefont {House}(2008)}]{house2008}%
  \BibitemOpen
  \bibfield  {author} {\bibinfo {author} {\bibfnamefont {M.~G.}\ \bibnamefont
  {House}},\ }\bibfield  {title} {\bibinfo {title} {Analytic model for
  electrostatic fields in surface-electrode ion traps},\ }\href
  {https://doi.org/10.1103/PhysRevA.78.033402} {\bibfield  {journal} {\bibinfo
  {journal} {Phys. Rev. A}\ }\textbf {\bibinfo {volume} {78}},\ \bibinfo
  {pages} {033402} (\bibinfo {year} {2008})}\BibitemShut {NoStop}%
\bibitem [{\citenamefont {Hahn}\ \emph
  {et~al.}(2019{\natexlab{b}})\citenamefont {Hahn}, \citenamefont
  {Zarantonello}, \citenamefont {Bautista-Salvador}, \citenamefont
  {Wahnschaffe}, \citenamefont {Kohnen}, \citenamefont {Schoebel},
  \citenamefont {Schmidt},\ and\ \citenamefont
  {Ospelkaus}}]{hahn2019multilayer}%
  \BibitemOpen
  \bibfield  {author} {\bibinfo {author} {\bibfnamefont {H.}~\bibnamefont
  {Hahn}}, \bibinfo {author} {\bibfnamefont {G.}~\bibnamefont {Zarantonello}},
  \bibinfo {author} {\bibfnamefont {A.}~\bibnamefont {Bautista-Salvador}},
  \bibinfo {author} {\bibfnamefont {M.}~\bibnamefont {Wahnschaffe}}, \bibinfo
  {author} {\bibfnamefont {M.}~\bibnamefont {Kohnen}}, \bibinfo {author}
  {\bibfnamefont {J.}~\bibnamefont {Schoebel}}, \bibinfo {author}
  {\bibfnamefont {P.~O.}\ \bibnamefont {Schmidt}},\ and\ \bibinfo {author}
  {\bibfnamefont {C.}~\bibnamefont {Ospelkaus}},\ }\bibfield  {title} {\bibinfo
  {title} {Multilayer ion trap with three-dimensional microwave circuitry for
  scalable quantum logic applications},\ }\bibfield  {journal} {\bibinfo
  {journal} {Applied Physics B}\ }\textbf {\bibinfo {volume} {125}},\ \href
  {https://doi.org/10.1007/s00340-019-7265-1} {10.1007/s00340-019-7265-1}
  (\bibinfo {year} {2019}{\natexlab{b}})\BibitemShut {NoStop}%
\bibitem [{\citenamefont {Aude~Craik}\ \emph {et~al.}(2014)\citenamefont
  {Aude~Craik}, \citenamefont {Linke}, \citenamefont {Harty}, \citenamefont
  {Ballance}, \citenamefont {Lucas}, \citenamefont {Steane},\ and\
  \citenamefont {Allcock}}]{AudeCraik2014}%
  \BibitemOpen
  \bibfield  {author} {\bibinfo {author} {\bibfnamefont {D.~P.~L.}\
  \bibnamefont {Aude~Craik}}, \bibinfo {author} {\bibfnamefont {N.~M.}\
  \bibnamefont {Linke}}, \bibinfo {author} {\bibfnamefont {T.~P.}\ \bibnamefont
  {Harty}}, \bibinfo {author} {\bibfnamefont {C.~J.}\ \bibnamefont {Ballance}},
  \bibinfo {author} {\bibfnamefont {D.~M.}\ \bibnamefont {Lucas}}, \bibinfo
  {author} {\bibfnamefont {A.~M.}\ \bibnamefont {Steane}},\ and\ \bibinfo
  {author} {\bibfnamefont {D.~T.~C.}\ \bibnamefont {Allcock}},\ }\bibfield
  {title} {\bibinfo {title} {Microwave control electrodes for scalable,
  parallel, single-qubit operations in a surface-electrode ion trap},\ }\href
  {https://doi.org/10.1007/s00340-013-5716-7} {\bibfield  {journal} {\bibinfo
  {journal} {Applied Physics B}\ }\textbf {\bibinfo {volume} {114}},\ \bibinfo
  {pages} {3} (\bibinfo {year} {2014})}\BibitemShut {NoStop}%
\bibitem [{\citenamefont {Warring}\ \emph
  {et~al.}(2013{\natexlab{b}})\citenamefont {Warring}, \citenamefont
  {Ospelkaus}, \citenamefont {Colombe}, \citenamefont {Brown}, \citenamefont
  {Amini}, \citenamefont {Carsjens}, \citenamefont {Leibfried},\ and\
  \citenamefont {Wineland}}]{Warring2013Phase}%
  \BibitemOpen
  \bibfield  {author} {\bibinfo {author} {\bibfnamefont {U.}~\bibnamefont
  {Warring}}, \bibinfo {author} {\bibfnamefont {C.}~\bibnamefont {Ospelkaus}},
  \bibinfo {author} {\bibfnamefont {Y.}~\bibnamefont {Colombe}}, \bibinfo
  {author} {\bibfnamefont {K.~R.}\ \bibnamefont {Brown}}, \bibinfo {author}
  {\bibfnamefont {J.~M.}\ \bibnamefont {Amini}}, \bibinfo {author}
  {\bibfnamefont {M.}~\bibnamefont {Carsjens}}, \bibinfo {author}
  {\bibfnamefont {D.}~\bibnamefont {Leibfried}},\ and\ \bibinfo {author}
  {\bibfnamefont {D.~J.}\ \bibnamefont {Wineland}},\ }\bibfield  {title}
  {\bibinfo {title} {Techniques for microwave near-field quantum control of
  trapped ions},\ }\href {https://doi.org/10.1103/PhysRevA.87.013437}
  {\bibfield  {journal} {\bibinfo  {journal} {Phys. Rev. A}\ }\textbf {\bibinfo
  {volume} {87}},\ \bibinfo {pages} {013437} (\bibinfo {year}
  {2013}{\natexlab{b}})}\BibitemShut {NoStop}%
\bibitem [{\citenamefont {Brownnutt}\ \emph {et~al.}(2015)\citenamefont
  {Brownnutt}, \citenamefont {Kumph}, \citenamefont {Rabl},\ and\ \citenamefont
  {Blatt}}]{Brownnutt2015}%
  \BibitemOpen
  \bibfield  {author} {\bibinfo {author} {\bibfnamefont {M.}~\bibnamefont
  {Brownnutt}}, \bibinfo {author} {\bibfnamefont {M.}~\bibnamefont {Kumph}},
  \bibinfo {author} {\bibfnamefont {P.}~\bibnamefont {Rabl}},\ and\ \bibinfo
  {author} {\bibfnamefont {R.}~\bibnamefont {Blatt}},\ }\bibfield  {title}
  {\bibinfo {title} {Ion-trap measurements of electric-field noise near
  surfaces},\ }\href {https://doi.org/10.1103/RevModPhys.87.1419} {\bibfield
  {journal} {\bibinfo  {journal} {Rev. Mod. Phys.}\ }\textbf {\bibinfo {volume}
  {87}},\ \bibinfo {pages} {1419} (\bibinfo {year} {2015})}\BibitemShut
  {NoStop}%
\bibitem [{\citenamefont {Vittorini}(2013)}]{Vittorini2013}%
  \BibitemOpen
  \bibfield  {author} {\bibinfo {author} {\bibfnamefont {G.~D.}\ \bibnamefont
  {Vittorini}},\ }\emph {\bibinfo {title} {Stability of ion chains in a
  cryogenic surface-electrode ion trap}},\ \href
  {http://hdl.handle.net/1853/50239} {Ph.D. thesis},\ \bibinfo  {school}
  {Georgia Tech} (\bibinfo {year} {2013}),\ \bibinfo {note} {section
  4.2}\BibitemShut {NoStop}%
\bibitem [{\citenamefont {Zhu}\ \emph {et~al.}(2006)\citenamefont {Zhu},
  \citenamefont {Monroe},\ and\ \citenamefont {Duan}}]{ShiLiang2006}%
  \BibitemOpen
  \bibfield  {author} {\bibinfo {author} {\bibfnamefont {S.-L.}\ \bibnamefont
  {Zhu}}, \bibinfo {author} {\bibfnamefont {C.}~\bibnamefont {Monroe}},\ and\
  \bibinfo {author} {\bibfnamefont {L.-M.}\ \bibnamefont {Duan}},\ }\bibfield
  {title} {\bibinfo {title} {Arbitrary-speed quantum gates within large ion
  crystals through minimum control of laser beams},\ }\href
  {https://doi.org/10.1209/epl/i2005-10424-4} {\bibfield  {journal} {\bibinfo
  {journal} {Europhysics Letters}\ }\textbf {\bibinfo {volume} {73}},\ \bibinfo
  {pages} {485} (\bibinfo {year} {2006})}\BibitemShut {NoStop}%
\bibitem [{\citenamefont {Choi}\ \emph {et~al.}(2014)\citenamefont {Choi},
  \citenamefont {Debnath}, \citenamefont {Manning}, \citenamefont {Figgatt},
  \citenamefont {Gong}, \citenamefont {Duan},\ and\ \citenamefont
  {Monroe}}]{choi2014}%
  \BibitemOpen
  \bibfield  {author} {\bibinfo {author} {\bibfnamefont {T.}~\bibnamefont
  {Choi}}, \bibinfo {author} {\bibfnamefont {S.}~\bibnamefont {Debnath}},
  \bibinfo {author} {\bibfnamefont {T.~A.}\ \bibnamefont {Manning}}, \bibinfo
  {author} {\bibfnamefont {C.}~\bibnamefont {Figgatt}}, \bibinfo {author}
  {\bibfnamefont {Z.-X.}\ \bibnamefont {Gong}}, \bibinfo {author}
  {\bibfnamefont {L.-M.}\ \bibnamefont {Duan}},\ and\ \bibinfo {author}
  {\bibfnamefont {C.}~\bibnamefont {Monroe}},\ }\bibfield  {title} {\bibinfo
  {title} {Optimal quantum control of multimode couplings between trapped ion
  qubits for scalable entanglement},\ }\href
  {https://doi.org/10.1103/PhysRevLett.112.190502} {\bibfield  {journal}
  {\bibinfo  {journal} {Phys. Rev. Lett.}\ }\textbf {\bibinfo {volume} {112}},\
  \bibinfo {pages} {190502} (\bibinfo {year} {2014})}\BibitemShut {NoStop}%
\bibitem [{\citenamefont {Kranzl}\ \emph {et~al.}(2022)\citenamefont {Kranzl},
  \citenamefont {Joshi}, \citenamefont {Maier}, \citenamefont {Brydges},
  \citenamefont {Franke}, \citenamefont {Blatt},\ and\ \citenamefont
  {Roos}}]{Kranzl2022}%
  \BibitemOpen
  \bibfield  {author} {\bibinfo {author} {\bibfnamefont {F.}~\bibnamefont
  {Kranzl}}, \bibinfo {author} {\bibfnamefont {M.~K.}\ \bibnamefont {Joshi}},
  \bibinfo {author} {\bibfnamefont {C.}~\bibnamefont {Maier}}, \bibinfo
  {author} {\bibfnamefont {T.}~\bibnamefont {Brydges}}, \bibinfo {author}
  {\bibfnamefont {J.}~\bibnamefont {Franke}}, \bibinfo {author} {\bibfnamefont
  {R.}~\bibnamefont {Blatt}},\ and\ \bibinfo {author} {\bibfnamefont {C.~F.}\
  \bibnamefont {Roos}},\ }\bibfield  {title} {\bibinfo {title} {Controlling
  long ion strings for quantum simulation and precision measurements},\ }\href
  {https://doi.org/10.1103/PhysRevA.105.052426} {\bibfield  {journal} {\bibinfo
   {journal} {Phys. Rev. A}\ }\textbf {\bibinfo {volume} {105}},\ \bibinfo
  {pages} {052426} (\bibinfo {year} {2022})}\BibitemShut {NoStop}%
\bibitem [{\citenamefont {Knill}\ \emph {et~al.}(2008)\citenamefont {Knill},
  \citenamefont {Leibfried}, \citenamefont {Reichle}, \citenamefont {Britton},
  \citenamefont {Blakestad}, \citenamefont {Jost}, \citenamefont {Langer},
  \citenamefont {Ozeri}, \citenamefont {Seidelin},\ and\ \citenamefont
  {Wineland}}]{Knill2008RBM}%
  \BibitemOpen
  \bibfield  {author} {\bibinfo {author} {\bibfnamefont {E.}~\bibnamefont
  {Knill}}, \bibinfo {author} {\bibfnamefont {D.}~\bibnamefont {Leibfried}},
  \bibinfo {author} {\bibfnamefont {R.}~\bibnamefont {Reichle}}, \bibinfo
  {author} {\bibfnamefont {J.}~\bibnamefont {Britton}}, \bibinfo {author}
  {\bibfnamefont {R.~B.}\ \bibnamefont {Blakestad}}, \bibinfo {author}
  {\bibfnamefont {J.~D.}\ \bibnamefont {Jost}}, \bibinfo {author}
  {\bibfnamefont {C.}~\bibnamefont {Langer}}, \bibinfo {author} {\bibfnamefont
  {R.}~\bibnamefont {Ozeri}}, \bibinfo {author} {\bibfnamefont
  {S.}~\bibnamefont {Seidelin}},\ and\ \bibinfo {author} {\bibfnamefont
  {D.~J.}\ \bibnamefont {Wineland}},\ }\bibfield  {title} {\bibinfo {title}
  {Randomized benchmarking of quantum gates},\ }\href
  {https://doi.org/10.1103/PhysRevA.77.012307} {\bibfield  {journal} {\bibinfo
  {journal} {Phys. Rev. A}\ }\textbf {\bibinfo {volume} {77}},\ \bibinfo
  {pages} {012307} (\bibinfo {year} {2008})}\BibitemShut {NoStop}%
\bibitem [{\citenamefont {Cetina}\ \emph {et~al.}(2022)\citenamefont {Cetina},
  \citenamefont {Egan}, \citenamefont {Noel}, \citenamefont {Goldman},
  \citenamefont {Biswas}, \citenamefont {Risinger}, \citenamefont {Zhu},\ and\
  \citenamefont {Monroe}}]{Cetina2022}%
  \BibitemOpen
  \bibfield  {author} {\bibinfo {author} {\bibfnamefont {M.}~\bibnamefont
  {Cetina}}, \bibinfo {author} {\bibfnamefont {L.}~\bibnamefont {Egan}},
  \bibinfo {author} {\bibfnamefont {C.}~\bibnamefont {Noel}}, \bibinfo {author}
  {\bibfnamefont {M.}~\bibnamefont {Goldman}}, \bibinfo {author} {\bibfnamefont
  {D.}~\bibnamefont {Biswas}}, \bibinfo {author} {\bibfnamefont
  {A.}~\bibnamefont {Risinger}}, \bibinfo {author} {\bibfnamefont
  {D.}~\bibnamefont {Zhu}},\ and\ \bibinfo {author} {\bibfnamefont
  {C.}~\bibnamefont {Monroe}},\ }\bibfield  {title} {\bibinfo {title} {Control
  of transverse motion for quantum gates on individually addressed atomic
  qubits},\ }\href {https://doi.org/10.1103/PRXQuantum.3.010334} {\bibfield
  {journal} {\bibinfo  {journal} {PRX Quantum}\ }\textbf {\bibinfo {volume}
  {3}},\ \bibinfo {pages} {010334} (\bibinfo {year} {2022})}\BibitemShut
  {NoStop}%
\bibitem [{\citenamefont {Landsman}\ \emph {et~al.}(2019)\citenamefont
  {Landsman}, \citenamefont {Wu}, \citenamefont {Leung}, \citenamefont {Zhu},
  \citenamefont {Linke}, \citenamefont {Brown}, \citenamefont {Duan},\ and\
  \citenamefont {Monroe}}]{Landsman2019}%
  \BibitemOpen
  \bibfield  {author} {\bibinfo {author} {\bibfnamefont {K.~A.}\ \bibnamefont
  {Landsman}}, \bibinfo {author} {\bibfnamefont {Y.}~\bibnamefont {Wu}},
  \bibinfo {author} {\bibfnamefont {P.~H.}\ \bibnamefont {Leung}}, \bibinfo
  {author} {\bibfnamefont {D.}~\bibnamefont {Zhu}}, \bibinfo {author}
  {\bibfnamefont {N.~M.}\ \bibnamefont {Linke}}, \bibinfo {author}
  {\bibfnamefont {K.~R.}\ \bibnamefont {Brown}}, \bibinfo {author}
  {\bibfnamefont {L.}~\bibnamefont {Duan}},\ and\ \bibinfo {author}
  {\bibfnamefont {C.}~\bibnamefont {Monroe}},\ }\bibfield  {title} {\bibinfo
  {title} {Two-qubit entangling gates within arbitrarily long chains of trapped
  ions},\ }\href {https://doi.org/10.1103/PhysRevA.100.022332} {\bibfield
  {journal} {\bibinfo  {journal} {Phys. Rev. A}\ }\textbf {\bibinfo {volume}
  {100}},\ \bibinfo {pages} {022332} (\bibinfo {year} {2019})}\BibitemShut
  {NoStop}%
\bibitem [{\citenamefont {Kielpinski}\ \emph {et~al.}(2002)\citenamefont
  {Kielpinski}, \citenamefont {Monroe},\ and\ \citenamefont
  {Wineland}}]{Wineland2002QCCD}%
  \BibitemOpen
  \bibfield  {author} {\bibinfo {author} {\bibfnamefont {D.}~\bibnamefont
  {Kielpinski}}, \bibinfo {author} {\bibfnamefont {C.}~\bibnamefont {Monroe}},\
  and\ \bibinfo {author} {\bibfnamefont {D.~J.}\ \bibnamefont {Wineland}},\
  }\bibfield  {title} {\bibinfo {title} {Architecture for a large-scale
  ion-trap quantum computer},\ }\href {https://doi.org/10.1038/nature00784}
  {\bibfield  {journal} {\bibinfo  {journal} {Nature}\ }\textbf {\bibinfo
  {volume} {417}},\ \bibinfo {pages} {709} (\bibinfo {year}
  {2002})}\BibitemShut {NoStop}%
\bibitem [{\citenamefont {Monroe}\ \emph {et~al.}(2014)\citenamefont {Monroe},
  \citenamefont {Raussendorf}, \citenamefont {Ruthven}, \citenamefont {Brown},
  \citenamefont {Maunz}, \citenamefont {Duan},\ and\ \citenamefont
  {Kim}}]{Monroe2014Networking}%
  \BibitemOpen
  \bibfield  {author} {\bibinfo {author} {\bibfnamefont {C.}~\bibnamefont
  {Monroe}}, \bibinfo {author} {\bibfnamefont {R.}~\bibnamefont {Raussendorf}},
  \bibinfo {author} {\bibfnamefont {A.}~\bibnamefont {Ruthven}}, \bibinfo
  {author} {\bibfnamefont {K.~R.}\ \bibnamefont {Brown}}, \bibinfo {author}
  {\bibfnamefont {P.}~\bibnamefont {Maunz}}, \bibinfo {author} {\bibfnamefont
  {L.-M.}\ \bibnamefont {Duan}},\ and\ \bibinfo {author} {\bibfnamefont
  {J.}~\bibnamefont {Kim}},\ }\bibfield  {title} {\bibinfo {title} {Large-scale
  modular quantum-computer architecture with atomic memory and photonic
  interconnects},\ }\href {https://doi.org/10.1103/PhysRevA.89.022317}
  {\bibfield  {journal} {\bibinfo  {journal} {Phys. Rev. A}\ }\textbf {\bibinfo
  {volume} {89}},\ \bibinfo {pages} {022317} (\bibinfo {year}
  {2014})}\BibitemShut {NoStop}%
\bibitem [{\citenamefont {Kr{\"a}mer}\ \emph {et~al.}(2018)\citenamefont
  {Kr{\"a}mer}, \citenamefont {Plankensteiner}, \citenamefont {Ostermann},\
  and\ \citenamefont {Ritsch}}]{kramer2018quantumoptics}%
  \BibitemOpen
  \bibfield  {author} {\bibinfo {author} {\bibfnamefont {S.}~\bibnamefont
  {Kr{\"a}mer}}, \bibinfo {author} {\bibfnamefont {D.}~\bibnamefont
  {Plankensteiner}}, \bibinfo {author} {\bibfnamefont {L.}~\bibnamefont
  {Ostermann}},\ and\ \bibinfo {author} {\bibfnamefont {H.}~\bibnamefont
  {Ritsch}},\ }\bibfield  {title} {\bibinfo {title} {Quantumoptics. jl: A julia
  framework for simulating open quantum systems},\ }\href
  {https://doi.org/10.1016/j.cpc.2018.02.004} {\bibfield  {journal} {\bibinfo
  {journal} {Computer Physics Communications}\ }\textbf {\bibinfo {volume}
  {227}},\ \bibinfo {pages} {109} (\bibinfo {year} {2018})}\BibitemShut
  {NoStop}%
\bibitem [{\citenamefont {Mogensen}\ and\ \citenamefont
  {Riseth}(2018)}]{Mogensen2018optim}%
  \BibitemOpen
  \bibfield  {author} {\bibinfo {author} {\bibfnamefont {P.~K.}\ \bibnamefont
  {Mogensen}}\ and\ \bibinfo {author} {\bibfnamefont {A.~N.}\ \bibnamefont
  {Riseth}},\ }\bibfield  {title} {\bibinfo {title} {Optim: A mathematical
  optimization package for julia},\ }\href
  {https://doi.org/10.21105/joss.00615} {\bibfield  {journal} {\bibinfo
  {journal} {Journal of Open Source Software}\ }\textbf {\bibinfo {volume}
  {3}},\ \bibinfo {pages} {615} (\bibinfo {year} {2018})}\BibitemShut {NoStop}%
\bibitem [{\citenamefont {Magnus}(1954)}]{Magnus}%
  \BibitemOpen
  \bibfield  {author} {\bibinfo {author} {\bibfnamefont {W.}~\bibnamefont
  {Magnus}},\ }\bibfield  {title} {\bibinfo {title} {On the exponential
  solution of differential equations for a linear operator},\ }\href
  {https://doi.org/https://doi.org/10.1002/cpa.3160070404} {\bibfield
  {journal} {\bibinfo  {journal} {Communications on Pure and Applied
  Mathematics}\ }\textbf {\bibinfo {volume} {7}},\ \bibinfo {pages} {649}
  (\bibinfo {year} {1954})}\BibitemShut {NoStop}%
\bibitem [{\citenamefont {James}(1998)}]{James1998}%
  \BibitemOpen
  \bibfield  {author} {\bibinfo {author} {\bibfnamefont {D.~F.~V.}\
  \bibnamefont {James}},\ }\bibfield  {title} {\bibinfo {title} {Quantum
  dynamics of cold trapped ions with application to quantum computation},\
  }\href {https://doi.org/10.1007/s003400050373} {\bibfield  {journal}
  {\bibinfo  {journal} {Applied Physics B}\ }\textbf {\bibinfo {volume} {66}},\
  \bibinfo {pages} {181} (\bibinfo {year} {1998})}\BibitemShut {NoStop}%
\bibitem [{\citenamefont {Weber}\ \emph
  {et~al.}(2024{\natexlab{b}})\citenamefont {Weber}, \citenamefont {Gely},
  \citenamefont {Hanley}, \citenamefont {Harty}, \citenamefont {Leu},
  \citenamefont {Löschnauer}, \citenamefont {Nadlinger},\ and\ \citenamefont
  {Lucas}}]{weber2024robust}%
  \BibitemOpen
  \bibfield  {author} {\bibinfo {author} {\bibfnamefont {M.~A.}\ \bibnamefont
  {Weber}}, \bibinfo {author} {\bibfnamefont {M.~F.}\ \bibnamefont {Gely}},
  \bibinfo {author} {\bibfnamefont {R.~K.}\ \bibnamefont {Hanley}}, \bibinfo
  {author} {\bibfnamefont {T.~P.}\ \bibnamefont {Harty}}, \bibinfo {author}
  {\bibfnamefont {A.~D.}\ \bibnamefont {Leu}}, \bibinfo {author} {\bibfnamefont
  {C.~M.}\ \bibnamefont {Löschnauer}}, \bibinfo {author} {\bibfnamefont
  {D.~P.}\ \bibnamefont {Nadlinger}},\ and\ \bibinfo {author} {\bibfnamefont
  {D.~M.}\ \bibnamefont {Lucas}},\ }\href@noop {} {\bibinfo {title} {Robust and
  fast microwave-driven quantum logic for trapped-ion qubits}} (\bibinfo {year}
  {2024}{\natexlab{b}}),\ \Eprint {https://arxiv.org/abs/2402.12955}
  {arXiv:2402.12955 [quant-ph]} \BibitemShut {NoStop}%
\end{thebibliography}%

\FloatBarrier
\clearpage
\onecolumngrid
\begin{center}
{\Large \textbf{Supplementary information}}
\end{center}
\makeatletter
   \renewcommand\l@section{\@dottedtocline{2}{1.5em}{2em}}
   \renewcommand\l@subsection{\@dottedtocline{2}{3.5em}{2em}}
   \renewcommand\l@subsubsection{\@dottedtocline{2}{5.5em}{2em}}
\makeatother
\let\addcontentsline\oldaddcontentsline

\renewcommand{\thesection}{\arabic{section}}

\twocolumngrid

\let\oldaddcontentsline\addcontentsline
\renewcommand{\addcontentsline}[3]{}
\let\addcontentsline\oldaddcontentsline
\renewcommand{\theequation}{S\arabic{equation}}
\renewcommand{\thefigure}{S\arabic{figure}}
\renewcommand{\thetable}{S\arabic{table}}
\renewcommand{\thesection}{S\arabic{section}}
\setcounter{figure}{0}
\setcounter{equation}{0}
\setcounter{section}{0}

\newcolumntype{C}[1]{>{\centering\arraybackslash}p{#1}}
\newcolumntype{L}[1]{>{\raggedright\arraybackslash}p{#1}}

\FloatBarrier

\section{Hamiltonian derivation and numerical approach}\label{sec:theory}
We first make the assumption that motional modes (other than the in-plane rocking mode) and internal ion states (other than the qubit states) play ``spectator'' roles and are not involved in the observed physics (justification in Sec.~\ref{sec:ramping}).
Since we make use of a ``$\pi$'' qubit which is only sensitive to linearly polarised magnetic fields, in the direction of the quantisation axis ``$x$'', we are interested in the interaction between the magnetic field component $B_x$ (expressed here as a phasor) and the ion.
Coupling to the ion's harmonic motion (oriented along a direction $\underline u = q_x\underline x +q_y\underline y $) arises from the changing field amplitude and phase as the ion moves by a distance $u$ with respect to its equilibrium position $u=0$
\begin{equation}
    \begin{aligned}
    B_x(u) &= B_x(0) + u[\partial_u B_x]|_{u=0} \\
    &= B_x(0) + u\left[q_x\partial_x B_x + q_y \partial_y B_x\right]|_{u=0}\ .
\end{aligned}
\label{eq:B_taylor_expansion}
\end{equation}
Under these assumptions, the interaction-picture Hamiltonian (in the rotating wave approximation) describing the interplay of microwave field, qubit states and ion motion reads~\cite{ospelkaus2008}
\begin{equation}
    \begin{aligned}
   \hat H =& \frac{\hbar\Omega}{2} e^{-i\left(\omega -\omega_q\right)t}\hat\sigma_+ \\
   &\otimes\left(e^{i\phi_\text{field}}+\eta e^{i\phi_\text{grad}}\left(e^{-i\omega_m t}\hat a+e^{i\omega_m t}\hat a^\dagger\right)\right)+\text{h.c.}
\end{aligned}
\end{equation}
where:
\begin{itemize}
    \item $\omega$, $\omega_q$, $\omega_m$ are the frequencies of the microwave field, qubit transition, and ion motion respectively
    \item $\hat \sigma_+ = \ket{1}\bra{0}$ acts on the qubit states $\ket{0},\ket{1}$
    \item $\hat a$ is the annihilation operator for the ion's harmonic motion, related to the ion's displacement operator $\hat u = u_\text{zpf}(\hat a+\hat a^\dagger)$ through the zero-point fluctuations in motion $u_\text{zpf}=\sqrt{\hbar/(2M\omega_m)}$ ($M$ is the mass of the ion)
    \item $\phi_\text{field}$, $\phi_\text{grad}$ are the phase of the field $B_x$ and its gradient $\partial_uB_x$ respectively
    \item $\Omega$ is the Rabi-frequency, quantifying the strength of the magnetic field and the magnetic dipole moment $\mu_x$ of the qubit transition $\hbar\Omega = \mu_x |B_x|$
    \item $\eta$ is the effective Lamb-Dicke parameter, quantifying the relative change in field resulting from zero-point fluctuations $\eta = u_\text{zpf}|\partial_uB_x|/|B_x|$.
\end{itemize}

For a single microwave tone on resonance with the qubit ($\omega=\omega_q$) we can neglect the motional coupling under the assumption $\eta\Omega\ll\omega_m$ and write the dynamical decoupling (DD) driving Hamiltonian shown in Eq.~(3).
The state-dependent force (SDF) arises in the case where the red and blue motional sidebands are driven with two microwave tones at frequencies $\omega_q-(\omega_m+\delta)$ and $\omega_q+(\omega_m+\delta)$ respectively.
Under the assumption $|\delta|\ll\omega_m\ll\omega_q$, we first discard terms rotating at the qubit frequency to obtain
\begin{equation}
    \begin{aligned}
    \hat H_{\text{SB}} &= \hat H_{\text{off}} + \hat H_{\text{SDF}}\\
    \hat H_{\text{SDF}} &= \frac{\hbar\eta\Omega}{2}  \hat{\sigma}_{+} \left(  e^{i\delta t} \hat{a} +  e^{-i\delta t} \hat{a}^{\dagger} \right) + \text{h.c.}\\
    \hat H_{\text{off}} &= \frac{\hbar\Omega}{2} e^{i (\phi_{\text{field}}-\phi_{\text{grad}})} \hat{\sigma}_{+} \left( e^{i\left( \omega_m + \delta\right)t} + e^{-i\left( \omega_m + \delta\right)t}\right) + \text{h.c.} \
    \end{aligned}
    \label{eq:supp_full_hamiltonian}
\end{equation}
The resulting state-dependent force Hamiltonian $\hat H_{\text{SDF}}$ was presented in the main text (Eq. 1) using the sideband interaction strength $\Omega_{\text{SB}} = \eta\Omega$ and $\delta=0$.
The second part of the Hamiltonian, $\hat H_{\text{off}}$, representing off-resonant carrier driving, is better rewritten as
\begin{equation} \label{eq18}
    \begin{aligned}
    \hat{H}_{\text{off}} &= \hbar \Omega e^{i (\phi_{\text{field}}-\phi_{\text{grad}})} \hat{\sigma}_{+} \cos \left( \left( \omega_m + \delta \right) t \right) + \text{h.c.} \ .\\
    \end{aligned}
\end{equation}
It drives small oscillations on the Bloch sphere (as long as $\Omega\ll\omega_m+\delta$) and has a negligible effect (see Sec.~\ref{sec_off_resonant_car}).

In simulations, we take into account the interaction of the motional mode's heating rate $\dot{\bar{n}}$ and initial thermal occupation $\bar n$.
The quantum dynamics of this open quantum system is captured by the Lindblad master equation
\begin{equation} \label{eq:Lindblad}
    \begin{aligned}
    \dot{\hat{\rho}} = &-\frac{i}{\hbar} \left[ \hat{H}_\text{SDF} +\hat{H}_\text{off} +\hat{H}_\text{DD} , \hat{\rho} \right]\\
    & + \frac{\dot{\bar{n}}}{2} \left( 2 \hat{a} \hat{\rho} \hat{a}^{\dagger} - \hat{a}^{\dagger} \hat{a} \hat{\rho} - \hat{\rho} \hat{a}^{\dagger} \hat{a} \right)\\
     & + \frac{\dot{\bar{n}}}{2}  \left( 2 \hat{a}^{\dagger} \hat{\rho} \hat{a} - \hat{a} \hat{a}^{\dagger} \hat{\rho} - \hat{\rho} \hat{a} \hat{a}^{\dagger} \right) \ ,
    \end{aligned}
\end{equation}
where the initial state of the motional mode is a thermal state with average occupation $\langle \hat n \rangle = \bar n $.
Numerical simulation of this master equation is carried out using the Julia QuantumOptics package \cite{kramer2018quantumoptics}.
Most of the parameters used in the simulations are dependent on the displacement of the ion with respect to the RF-null of the trap (see Sec.~\ref{sec:experimental_considerations}).

\section{Microwave field and Biot-Savart model} \label{sec:two_wire_model}

\begin{figure} [t]
    \centering
    \includegraphics[width=0.45\textwidth]{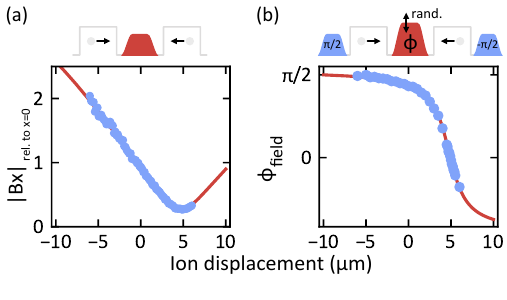}
    \caption{
    \textbf{Microwave field amplitude and phase.}
    \textbf{(a)} Measured microwave field amplitude (blue) compared to a fitted Biot-Savart model (red), relative to the amplitude at 0 displacement.
    The amplitude is measured by preparing (and measuring) the $\ket{0}$  state at x = 0 before (and after) displacing the ion to position x, where a MW pulse of fixed amplitude and duration is applied.
    \textbf{(b)} Microwave field phase.
    The phase is measured by preparing (and measuring) the $\ket{+}$ state at $x=0$ before (and after) displacing the ion to position $x$, where it is subject to a MW pulse with varying phase $\phi_p$ and a random amplitude.
    When the phase of this MW pulse is in the same direction as the state $\ket{+}$, the pulse has no effect on the state of the ion, and so a $-\pi/2$ pulse will return the ion to state $\ket{0}$.
    However, when the phase of this MW pulse of random amplitude is in a different direction, the pulse will drive some rotation such that a $-\pi/2$ pulse will not return the ion to its initial state.
    Averaging over multiple random pulse amplitudes ensures that the pulse of phase $\phi_p$ does not accidentally drive a 2$\pi$ rotation.
    For each ion displacement, we sweep the phase $\phi_p$ to maximise the probability of measuring $\ket{0}$ at the end of the pulse sequence.
    This constitutes a measurement of the field's position dependent phase shift.
    }
    \label{figS:bfield_map}
\end{figure}

The microwave fields generated by the electrodes in the measured trap and in the proposed three-electrode trap of Fig. 5 are calculated using the Biot-Savart law.
The underlying assumption is that, since the ion-electrode distance ($40\ \upmu$m) exceeds the dimensions of the electrode ($5\ \upmu$m height and $8\ \upmu$m width), the electrode is well approximated by a 1D wire.
The total field is computed by adding the magnetic fields produced by two of these effective wires, corresponding to currents traversing the microwave electrode on each side of the ion.
We fit this model to the measured change in microwave-field amplitude and phase shown in Fig.~\ref{figS:bfield_map}.

The free parameters in this fit are the relative phase and amplitude in the currents traversing the two wires.
These parameters are determined (in an idealised microwave electrode) by the losses and boundary condition of the electrode respectively~\cite{weber2022cryogenic}.
In our experiment, we find that we cannot predict the fitted values from the length and qualify factor of the microwave electrode (resonator).
As is covered in detail in Ref.~\cite{weber2022cryogenic}, coupling of the microwaves to the RF electrode plays an important role.
Regardless of these issues, as shown in Fig.~\ref{figS:bfield_map}, the Biot-Savart model with two free parameters leads to excellent fits to the measured microwave field.
We determine the relative phase to be 0.0480 radians and the relative amplitude to be 0.856.
We therefore take the same approach in computing the field for the three electrode case of Fig. 5.

\begin{figure}[t]
    \centering
    \includegraphics[width=0.45\textwidth]{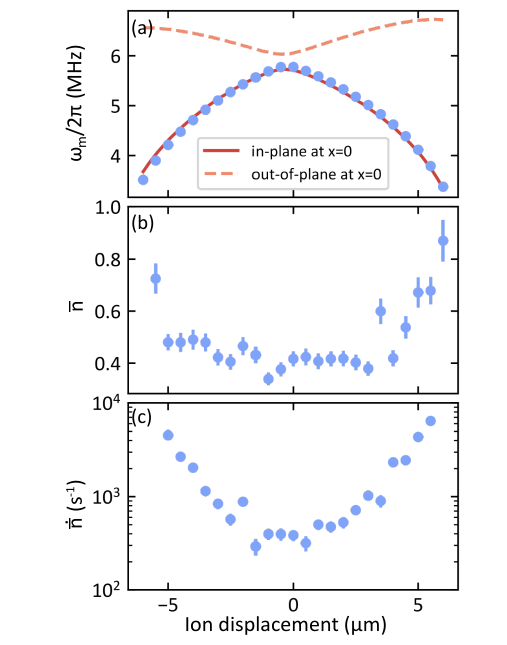}
    \caption{
    \textbf{Calibrations at displaced ion positions.}
    \textbf{(a)} Measured in-plane mode secular frequency (light blue) compared to simulated values for the in-plane mode (red) and out-of-plane mode (orange).
    Measured motional mode occupation \textbf{(b)} and heating rate \textbf{(c)} after sideband cooling at x=0 and displacing the ion to position x.
    We expect the increase in $\bar{n}$ is imposed by the changing $\dot{\bar{n}}$ at displaced positions, and motion induced by the displacement action.
    We attribute the increased $\dot{\bar{n}}$ to moving the ion away from the RF null, where fluctuations in the RF amplitude lead to a fluctuating force on the ion.
    }
    \label{figS:calibrations}
\end{figure}

\section{Displaced-ion calibrations} \label{sec:experimental_considerations}

\begin{figure} [t]
    \centering
    \includegraphics[width=0.45\textwidth]{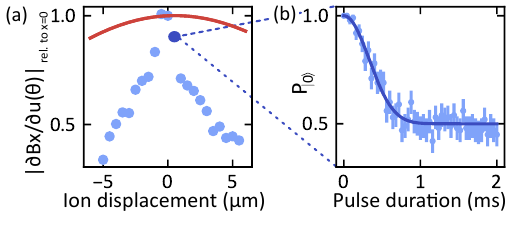}
    \caption{
    \textbf{Amplitude of the gradient of the microwave field.}
    \textbf{(a)} Measured amplitude of the gradient of the microwave field (blue) relative to the amplitude at zero displacement.
    This is compared to the predicted amplitude from the Biot-Savart model, not accounting for motional mode tilting (red).
    \textbf{(b)} Fitted simulation of a sideband pulse duration scan used to determine the microwave gradient, based on independently-determined mode properties (heating rate, thermal occupation, frequency).
    }
    \label{figS:mode_tilt_meas}
\end{figure}

\begin{figure} [t]
    \centering
    \includegraphics[width=0.45\textwidth]{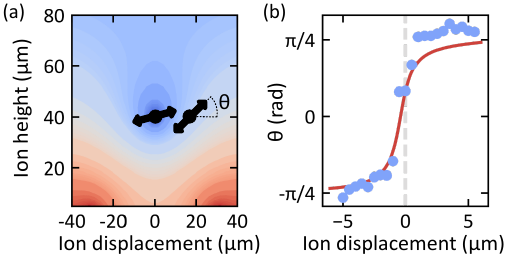}
    \caption{
    \textbf{Estimated motional mode tilt.}
    \textbf{(a)} Simulated pseudopotential generated by RF trapping fields in our surface trap.
    At zero displacement, the DC electrode voltages are chosen such that the motional mode tilt angle, $\theta$, is 15$^{\circ}$.
    As the ion is displaced, the curvature of the potential changes, causing the direction of the radial motional modes to tilt.
    \textbf{(b)} Tilt angle of the in-plane mode (light blue) extracted from the measured MW field gradient of Fig.~\ref{figS:mode_tilt_meas}(a) and compared to the expected tilt angle predicted from the simulated pseudopoential (red).
    }
    \label{figS:mode_tilt_5wire_check}
\end{figure}

\begin{figure} [t]
    \centering
    \includegraphics[width=0.45\textwidth]{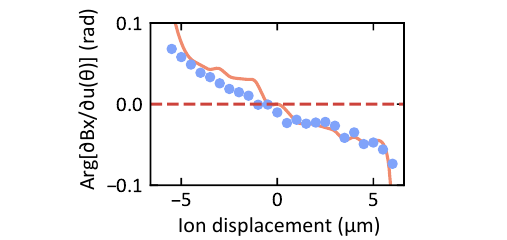}
    \caption{
    \textbf{Phase of the gradient of the microwave field.}
    Measured phase of the gradient of the microwave field (blue) compared to a Biot-Savart model when neglecting (red dashed line) or accounting for (orange solid line) the motional mode tilt.
    The phase of the gradient is measured relative to the field by maximising P$_{\ket{0}}$ in a DD phase scan, as described in Fig. 2(c).
    The phase of the gradient alone is then extracted using the independently measured phase of the field [Fig.~\ref{figS:bfield_map}(b)].
    }
    \label{figS:mode_tilt_gradphase_check}
\end{figure}

The ion is displaced away from the RF null in the x direction by changing the voltage applied to the DC electrodes, which has the undesired effects of changing the radial motional mode frequency $\omega_m$, the axis of motion (characterised by the tilt angle $\theta$), and the heating rate $\dot{\bar n}$, as well as introducing an additional heating mechanism (we move the ion after its sideband cooling sequence) which modifies the initial thermal population $\bar n$.
Here we explain how we disentangle these effects from the change in microwave field to best illustrate the ideas presented in Figs. 1,5.
We first perform calibrations of the mode frequency, heating and thermal occupations at different ion displacements from the RF null.
The resulting calibrations are shown in Fig.~\ref{figS:calibrations}.
The motional mode tilt angle $\theta$ (which we cannot straightforwardly measure) is then the only free parameter in fitting the decay of $P_{\ket{0}}$, assuming that the microwave field is well described by the Biot-Savart model.
Indeed, a change in $\theta$ changes the axis $\underline{u}$ along which the microwave gradient should be computed (Eq.~\ref{eq:B_taylor_expansion}), in turn changing the sideband interaction strength.
The angle $\theta$ is thus determined by fitting a pulse duration scan, as shown for example in Fig.~\ref{figS:mode_tilt_meas}(b), with a simulation of the Lindblad master equation~\ref{eq:Lindblad} using a least-squares minimisation routine~\cite{Mogensen2018optim}.
The resulting field gradient extracted at different positions is shown Fig.~\ref{figS:mode_tilt_meas}(a).
This approach relies on the accuracy of the ``Biot-Savart'' model.
One way in which we test this assumption is to check whether the fitted change in tilt angle is consistent with the expected pseudopoential at the displaced ion positions.
By making the assumption that these electrodes have infinitely small gaps between them, we can analytically estimate the pseudopoential governing the ion's harmonic motion~\cite{house2008}.
The RF contribution to the pseudopotential is shown in Fig.~\ref{figS:mode_tilt_5wire_check}(a).
The direction of motion is then given by the local curvature of this pseudopotential, which shows good agreement with the value independently extracted through measurements of the SDF, see Fig.~\ref{figS:mode_tilt_5wire_check}(b).
Note that there is an intentional $15^{\circ}$ tilt imposed at zero displacement, in order for the out-of-plane radial motion of the ion to have a small projection on the horizontally-delivered Doppler cooling laser beams~\cite{weber2022cryogenic}.
The second way in which we verify the extracted tilt angles is by comparing the measured phase of the microwave gradient -- which would be constant if $\theta=0$ -- to values predicted by the extracted tilt angle $\theta$.
The comparison is shown in Fig.~\ref{figS:mode_tilt_gradphase_check} and again demonstrates a good understanding of the change in motional tilt angle.

\section{Microwave pulse sequences}
\label{sec:microwave_pulse_sequence}

\subsection{Accounting for motional mode changes with displacement}

\begin{figure} [t]
    \centering

    \includegraphics[width=0.45\textwidth]{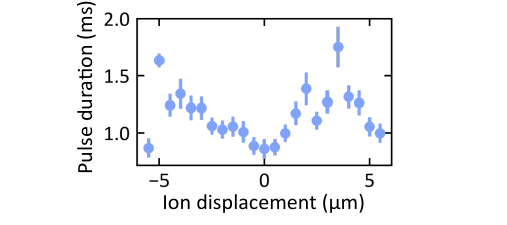}
    \caption{
    \textbf{Required pulse duration to maximally entangle spin and motion.}
    Sideband pulse duration required to reach P$_{\ket{0}} = $ 0.51 when driving the SDF at displaced ion positions.
    This is determined by fitting a Gaussian decay to a SB pulse duration scan (as in Fig. 2(b)).
    }
    \label{figS:pulse_duration}
\end{figure}

Keeping in mind our intention to demonstrate the ideas of Figs. 1,5, we vary the microwave drive parameters to simulate the driving of an ion chain with three electrodes.
In an ion chain, the motional mode is shared by all the ions, and therefore mode characteristics are independent of ion position.
We simulate this by changing the microwave drive frequencies as the ion is displaced from the RF null such that resonance with the motional sidebands is maintained.
Additionally, changes in motional occupation, heating rate, zero-point motion and motional tilt angle are disentangled from the effective SDF measurement shown in Fig. 4 by adjusting the microwave pulse duration.
The appropriate pulse duration for a given position is determined through a pulse duration scan.
After fitting the decay of $P_{\ket{0}}$ to a Gaussian (Eq. 6), we determined the pulse duration required to reach an excited state population of 0.51.
These values are shown for different ion positions in Fig.~\ref{figS:pulse_duration}.
By using this pulse duration, we maximise the change in $P_{\ket{0}}$ to best reveal the suppression of the effect of the SDF, as shown in Fig. 4(b).

\subsection{Dynamical decoupling drive amplitude and Walsh sequence selection} \label{sec:carrier_amplitude}

The DD drive should fulfill two conditions: (1) its strength should exceed that of the sideband interaction to suppress the effect of the SDF and (2) any rotations that it induces should be reversed by the end of the sequence.
In this section we explain how these two requirements are achieved through selection of an appropriate Walsh scheme \cite{Hayes2012Walsh}.

\begin{figure} [t]
    \centering
    \includegraphics[width=0.45\textwidth]{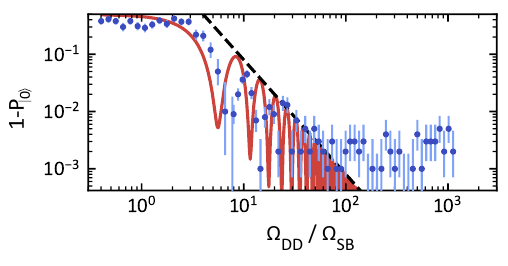}
    \caption{
        \textbf{Dynamical decoupling amplitude required to suppress the effect of the SDF.}
    For a Walsh-3 scheme, the DD strength $\Omega_\text{DD}$ is varied relative to a fixed sideband interaction strength $\Omega_\text{SB}$, illustrating that suppression occurs when the DD driving dominates $\Omega_\text{DD}\gg\Omega_\text{SB}$.
    This is compared to a simulation (red), where the maximum suppression error follows the empirically determined scaling of Eq.~(\ref{eq:carrier_strength_scaling}) (black dashed line).
    }
    \label{figS:carrier_amplitude}
\end{figure}

\begin{figure} [t]
    \centering
    \includegraphics[width=0.45\textwidth]{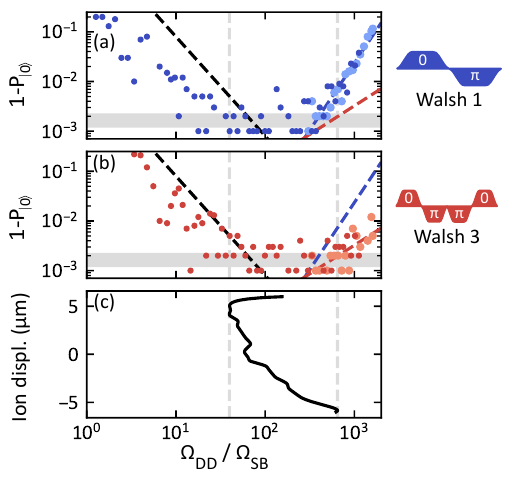}
    \caption{
    \textbf{Comparison of different Walsh sequences.}
    \textbf{(a/b)} Measured excited state population for varying DD strengths, when using a Walsh-1 (dark blue) or Walsh-3 (red) pulse scheme to cancel qubit rotations induced by DD.
    At lower DD strengths, the maximum measured suppression of the effect of the motional interaction follows the empirically determined scaling of Eq.~(\ref{eq:carrier_strength_scaling}) (black dashed line), as shown in Fig.~\ref{figS:carrier_amplitude}.
    At larger DD strengths, the measured suppression of the effect of the motional interaction is limited by incomplete cancellation of the effect of the DD.
    This was confirmed by detuning the sidebands 100 kHz from the secular frequency such that no motional interaction is driven (data: light blue/orange points, fits: blue/red dashed lines).
    The grey area shows the state preparation and measurement (SPAM) error.
    \textbf{(c)} Measured DD strength compared to sideband strength for displaced ion positions (deduced from the data shown in Fig.~\ref{figS:bfield_map} and Fig.~\ref{figS:mode_tilt_meas}).
    The light grey dashed lines indicate the minimum and maximum strengths in the range of ion displacements used for these experiments.
    This shows that we can sufficiently suppress the effect of the SDF -- to an error indistinguishable from the SPAM error by a straightforward measurement, see Fig.S11 for a more precise measurement of the suppression error --  for this entire range of DD/sideband strengths when using a Walsh-3 sequence.
    }
    \label{figS:decoupling_strength}
\end{figure}

The requirement for suppression, $\Omega_\text{DD}\gg\Omega_\text{SB}$, is illustrated in Fig.~\ref{figS:carrier_amplitude} for a Walsh-3 pulse sequence.
The suppression error oscillates with DD strength, presumably due to the frequency of the Rabi oscillations driven by the DD increasing with DD strength, while the pulse duration remains constant, such that the suppression is increased at some integer number of  $\Omega_\text{DD}/ \Omega_\text{SB}$.
Therefore, a fairer estimate of the expected suppression error is the maximum error for which, through simulations, we determine an empirical scaling
\begin{equation} \label{eq:carrier_strength_scaling}
    1 - P_{\ket{0}} = 8 \left(\frac{\Omega_{\text{DD}}}{\Omega_{\text{SB}}}\right)^{-2} .
\end{equation}
Mismatch between simulated and measured outcomes is expected due to an unknown transfer function distorting the DD and sideband amplitudes at the ion’s position, relative to the amplitudes at the signal generator.
We note that the strength of the sidebands, $\Omega_{\text{SB}}$, includes the effective Lamb-Dicke parameter $\eta$, which for our experiment is $1.25 \times 10^{-3}$.
This means that we can fulfill the condition $\Omega_\text{DD}\gg\Omega_\text{SB}$ using a lower power microwave current for the DD than the sidebands.
For stronger DD driving, we encounter the limits of the Walsh-3 pulse sequence, which we will now investigate in more detail.
A Walsh pulse scheme is used to ensure that no net rotation is driven by the DD by the end of the pulse sequence.
In this scheme each MW pulse is split into $2^N$ pulses of equal duration and the phases of half of these pulses are shifted by $\pi$ radians.
The choice of pulses which are $\pi$-shifted follows a Walsh sequence of order $2^N-1$.
A Walsh sequence of order 0 is the trivial case of a single pulse.
A Walsh sequence of order 1 (schematically shown in Fig.~\ref{figS:decoupling_strength}(a)) corresponds to two $\pi$-shifted pulses.
In this case any rotation accumulated in the first pulse will be cancelled in the second, assuming the DD drive strength remains constant.
A Walsh sequence of order 3 follows a more complex phase shifting scheme featuring four seperate pulses (schematically shown in Fig.~\ref{figS:decoupling_strength}(b)).
The advantage gained from this more complex pulse sequence is cancellation of rotations even if the DD drive strength varies linearly throughout the sequence.
A Walsh sequence of order 7 will cancel rotations in the presence of quadratic changes, order 15 cancels cubic changes, and so on.
Here, we compare the Walsh-1 scheme to the Walsh-3 scheme used in most other experiments, except the benchmarking experiment described in Sec.~\ref{sec:rbm}, which uses Walsh-15.

For varying amplitudes of DD driving, and different Walsh orders, we measured the change in initial state following a ``suppressed SDF'' pulse as described in Fig. 2(d).
In Fig.~\ref{figS:decoupling_strength}, we show that the Walsh-1 sequence is less effective at cancelling the net effect of the DD at high DD drive amplitudes than the Walsh-3 scheme.
This would suggest that the drive strength varies linearly with time throughout the sequence, for example due to thermal transients, making Walsh-3 more effective.
This is verified in Fig.~\ref{figS:decoupling_strength} by performing the same experiment but with the sideband frequencies symmetrically far-detuned from the motional sideband (by $2\pi\times$100 kHz), hence solely observing the consequences of DD driving.

As a result of the microwave field interference underpinning the DD phase shift, the amplitude of the DD drive will decrease with field amplitude, as shown in Fig.~\ref{figS:bfield_map}.
We chose however to keep the microwave power driving the DD transition constant, to best represent the situation that would be present in Fig. 5.
Furthermore, we select a DD drive amplitude which does not break the suppression condition $\Omega_\text{DD}\gg\Omega_\text{SB}$ as the ion is displaced away from the RF null.
Note that this decision makes our experiment a ``worst possible'' case, as a lower DD drive could be chosen to relax this condition at the interaction zone, minimising interference of the DD and gate interaction.
The range of relative DD amplitudes sampled as the ion is displaced is shown in Fig.~\ref{figS:decoupling_strength}(c).

\subsection{Impact of off-resonant carrier excitation} \label{sec_off_resonant_car}

Here we give support to the claim that off-resonant carrier excitations driven by sidebands $\hat H_\text{off}$ have negligible consequences.
The first impact of this term to consider is the net rotation it may drive on the qubit over the course of a pulse.
The second issue to discuss is that this term does not commute with the SDF in the interaction zones, and does not commute with the DD in the memory zones.

From Eq.~\ref{eq18}, ignoring phases, and for resonant sideband driving, the interaction Hamiltonian of interest is
\begin{equation} \label{off_resonant_eq}
    \hat{H}_{\text{off}} = \Omega \cos \left( \omega_m  t \right)\hat \sigma_x
\end{equation}
where $\Omega = \Omega_{\text{SB}}/\eta$ is the strength of the drive and $\omega_m$ is the detuning of the sidebands from the qubit frequency.
At the RF null, we have used $\Omega_{\text{SB}} = 2\pi\times$ 380 Hz and $\omega_m=2\pi\times$ 5.77 MHz.
Disregarding its commutation relation with other terms of the Hamiltonian, and assuming an initial qubit state $\ket{0}$, $\hat{H}_{\text{off}}$ will drive small oscillations of $\langle\hat\sigma\rangle$ with period $2\pi/\omega_m$ and amplitude $2(\Omega/\omega_m)^2$ when $\Omega\ll\omega_m$.
We ramp the sideband microwave amplitude following a $\sin^2$ shape over a duration of $2.4\ \upmu$s $\gg 2\pi/\omega_m$ to ensure that no net rotation is driven by the end of the gate.
We study the impact of the non-commutativity of $\hat{H}_{\text{off}}$ with other terms through simulation.
We restrict our analysis to the parameter regime used in this demonstration at the RF null.
For the memory zone, $\hat{H}_{\text{off}}$ does not commute with $\hat{H}_{\text{DD}}$, and therefore may limit its effectiveness.
This is already taken into account in the DD strength simulation of Fig.~\ref{figS:carrier_amplitude}, demonstrating that for sufficiently large DD drive strengths, there is no determined effect due to off-resonant (and out-of-phase) carrier driving.
For the interaction zone, we simulate the fidelity of a hypothetical single-loop two-qubit gate using the same sideband interaction and DD drive strengths as in our single-ion experiment at the RF-null.
The gate duration of such a gate would be $1.27$ ms, consistent with typical microwave gate durations~\cite{harty2016}.
The Walsh-3 scheme is applied to the DD drive in the simulated gate.
The impact on (Bell-state preparation) fidelity, even for off-resonant driving rates up to ten times larger than present in the current experiment, is negligible ($\lesssim 10^{-6}$).
Note that the strength of the off-resonant carrier excitation, for a given gate duration, will be inversely proportional to the effective Lamb-Dicke parameter $\eta$.
The fact that, in the scheme described in Fig. 5, gates are driven where the microwave field is suppressed by interference will increase $\eta$ and decrease the relevance of off-resonantly driven carrier excitations in interaction zones.

\subsection{Microwave pulse ramping scheme} \label{sec:ramping}

The microwave amplitude of both sideband and DD drive should be ramped up and down slowly to adiabatically transition in and out of hybridised states of:
\begin{enumerate}
    \item The qubit and out-of-plane motional spectator mode, driven with strength $\Omega /2\pi <380$ Hz by the sidebands detuned by $\Delta/2\pi\gtrsim$300 kHz
    \item The two qubit states, driven off-resonantly with strength $\Omega/2\pi \lesssim 640$ kHz by the sidebands detuned by $\Delta = \omega_m  \sim 2\pi\times$3--6 MHz
    \item The qubit and spectator hyperfine states, driven by both DD and sidebands detuned by $\Delta/2\pi\sim$100 MHz
\end{enumerate}
The impact of these off-resonant interactions on qubit state populations will be, following first order perturbation theory, of order $\sim(\Omega/\Delta)^2$.
This renders the impact of the out-of-plane mode negligible.
For the other two interactions, the DD and sideband ramping rate should far exceed the period of oscillations being driven ($\sim 1/\Delta$): $10$ ns and $300$ ns respectively.
Otherwise, these interactions would have an impact of order $10^{-4}$ and $10^{-2}$ respectively.
To satisfy this condition, each pulse of the Walsh modulation sequence starts and ends with a sin$^2$ shaped ramping period of durations $120$ ns and $2.4\ \upmu$s for the DD and sidebands respectively.
For effective suppression of the SDF, we always ramp up/down the DD driving before/after ramping up the sideband drive.

\subsection{Compensating for AC Zeeman shifts}

The microwave sidebands off-resonantly drive the qubit, as well as ``spectator'' hyperfine transitions, hybridising the qubit states and causing AC Zeeman shifts.
This scheme loses effectiveness around x $= 5\ \upmu$m, requiring a much larger imbalance, as the $\pi$-polarised field gets smaller (Fig.~\ref{figS:bfield_map}(a)) whereas the $\sigma^{-}$field undergoes little change.
A $\pi$-polarised (linearly polarised in the $x$ direction) magnetic field drives transitions where $\Delta M = 0$, such as the qubit transition.
Since the red and blue sidebands are oppositely detuned from the qubit frequency
($\omega_\text{r} = \omega_{q} - \delta_\text{r}$ and $\omega_\text{b} = \omega_{q} + \delta_\text{b}$)
the red sideband will increase the qubit frequency whilst the blue sideband will decrease the qubit frequency.
These contributions cancel with sidebands of equal amplitude, but a small imbalance in the sidebands can be used to compensate the other AC Zeeman shift.

\begin{figure} [t]
    \centering
    \includegraphics[width=0.45\textwidth]{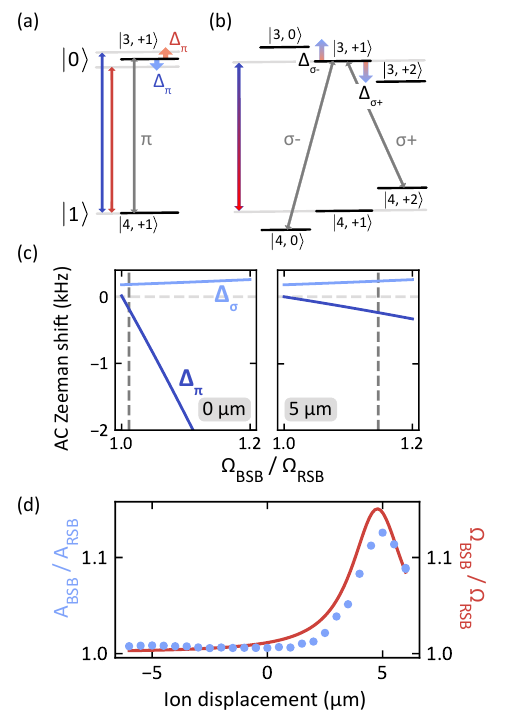}
    \caption{
    \textbf{Compensating for the AC Zeeman shift.}
    \textbf{(a)} Red and blue sidebands are oppositely detuned from the $\pi$-transition, creating AC Zeeman shifts in opposite directions.
    \textbf{(b)} Red and blue sidebands are both detuned from the $\sigma$-transitions in the same direction.
    However, both sidebands create $\sigma_{-}$ and $\sigma_{+}$ AC Zeeman shifts, which cancel themselves if the manifold is symmetric.
    The schematic shows the shift caused a single sideband on one energy level, but an identical shift will be produced both by the other sideband and by both sidebands on the ground state of the qubit.
    \textbf{(c)} Total shifts created by both sidebands for the $\sigma$ (light blue) and $\pi$ (dark blue) transitions, at ion displacements of x $= 0~\upmu\text{m}$ and x $= 5~\upmu\text{m}$.
    The amplitude of the blue sideband tone can be adjusted such that the two shifts cancel each other out (vertical dashed line).
    \textbf{(d)} Measured (blue) and expected (red) sideband amplitude imbalance required to compensate for the net AC Zeeman shift experienced by the ion.
    The expected imbalance is determined from the change in MW field amplitude (Fig.~\ref{figS:bfield_map}(a)).
    }
    \label{figS:ac_zeeman_shift}
\end{figure}

A $\sigma$-polarised magnetic field will cause frequency shifts of transitions where $\Delta M = \pm 1$, connecting the qubit states to states with $M = 0,2$ (see Fig.~\ref{figS:ac_zeeman_shift}(b)).
Contrary to the $\pi$-shift, both sidebands have a similar detuning to each off-resonantly driven transition, and will cause shifts in the same direction.
However, $\sigma_{-}$/$\sigma_{+}$ transitions connecting the qubit to lower and higher $M$ states are negatively/positively detuned from the sideband frequencies, creating shifts $\Delta_{\sigma_-}$/$\Delta_{\sigma_+}$ of opposite signs.

The $\sigma$-shifts do not completely cancel, resulting in a microwave-power dependent shift $\Delta_{\sigma} = \Delta_{\sigma_-}+\Delta_{\sigma_+}$.
This shift can be cancelled (for all amplitudes), by imposing a small sideband amplitude imbalance, which has a much larger impact on the magnitude of $\Delta_\pi$ than $\Delta_\sigma$, as shown in Fig.~\ref{figS:ac_zeeman_shift}(c).
This scheme loses effectiveness around x $= 5\ \upmu$m, requiring a much larger imbalance, as the $\pi$-polarised field gets smaller (Fig.~\ref{figS:bfield_map}(a)) whereas the $\sigma^{-}$field undergoes little change.
Whilst we use this approach to cancel AC Zeeman shifts, detuning microwave tones is an alternative solution, although it has the added complexity of amplitude dependence in the shift.

We calibrate the imbalance required to cancel all AC Zeeman shifts by applying the sidebands together with a weak tone at the qubit frequency, calibrated to map the initial state $\ket{0}$ to $\ket{1}$ when on resonance.
The required imbalance is determined by minimising $P_{\ket{0}}$ for different values of the imbalance.
The change in required imbalance is shown in Fig.~\ref{figS:ac_zeeman_shift}(d), and compared to the expected imbalance based on the change in MW field (Fig.~\ref{figS:bfield_map}(a)).
Mismatch to the theoretical value is expected due to an unknown and frequency-dependent transfer function distorting the RSB/BSB amplitude at the ion's position ($\Omega_\text{RSB,BSB}$), relative to the amplitude ($A_\text{RSB,BSB}$) at the signal generator.

\section{Ion chain control}

\subsection{Crosstalk estimation}
\label{sec:three_electrode_sims}

\begin{figure*}
    \centering
    \includegraphics[width=\textwidth]{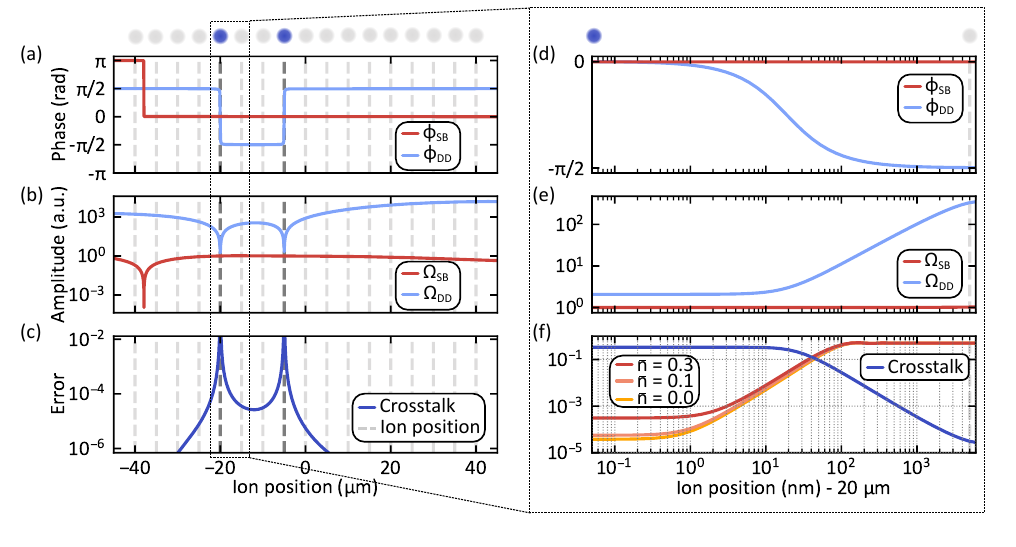}
    \caption{
    \textbf{Crosstalk simulation for a long ion chain.}
    \textbf{(a)} Using a simple three wire model, we use the Biot-Savart law to calculate the optimal fields and phases to address the ions at positions $-20~\upmu$m and $-5~\upmu$m.
    The phases of the DD (blue) and the sideband (red) are out of phase at all ion positions (light grey dashed lines) except at the addressed ions (dark grey dashed lines)
    \textbf{(b)} Interaction strength of the sideband (red) changes across the ion chain, optimised to have the same amplitude at the two addressed ion positions.
    The ratio of amplitudes $\Omega_\text{DD}/\Omega_{SB}$ guarantees $<7\times 10^{-5}$ crosstalk errors as given by Eq.~(\ref{eq:carrier_strength_scaling}).
    \textbf{(c)} Estimated crosstalk calculated as a purity loss arising from the residual entangling interaction.
    \textbf{(d-e)} Phase and Rabi frequency changes around the left-most addressed ion.
    \textbf{(f)} In addition to the crosstalk calculation, here we show the gate error induced by changes in the addressed ion position (shades of orange), for varying thermal occupation $\bar n$, induced by the decoupling tone on the addressed ion.
    }
    \label{figS:crosstalk_fields}
\end{figure*}

In this section, we discuss the assumptions and limitations of the crosstalk simulation presented in Fig. 5.
We consider a chain of 17 equally-spaced ions, with an ion-ion distance of $5~\upmu$m, located $40~\upmu$m above a surface trap.
These values were chosen to match the ion height and typical inter-ion-spacing used in our trap.
We envisage a geometry where, unlike our present trap, the MW electrodes are perpendicular to the ion string; this could be created using a multi-layer surface trap, where buried microwave electrodes are brought to the surface under the ions \cite{hahn2019multilayer}.
The microwave electrodes, $40~\upmu$m apart from each other, were simplified to wires to model the magnetic field using the Biot-Savart law; this was shown to provide reliable predictions for our electrode geometry (see Sec.~\ref{sec:two_wire_model}).
The three electrodes are assumed to be fed by independent microwave currents, where the central MW current is out-of-phase with the other two, but offset by $3\times 10^{-5}$ radians.
This value is based on experiments of Ref.~\cite{Warring2013Phase}, where phase control was sufficient to suppress the MW amplitude by a factor 40 with respect to an ion position 350 $\upmu$m away from the point of maximum interference.
Using the Biot-Savart model of Sec.~\ref{sec:two_wire_model}, we estimate the phase uncertainty demonstrated in Ref.~\cite{Warring2013Phase} to be $\sim 3\times 10^{-5}$ radians.
For every pair of ions, we use an optimisation routine (least-squares method) to determine the amplitudes of the sideband and DD currents required to address the selected ion-pair.
Firstly, the amplitudes of the sideband currents are chosen to minimise the difference in field gradient at the two addressed ion positions.
This ensures the state-dependent force will be equal at both positions during a two-qubit gate.
Secondly, the DD currents are chosen to minimise the phase-difference between the field and the gradient at the two addressed ion positions.
Using the optimum current configuration, we then compute the field and its gradient at all ion positions to determine the residual state-dependent force.
An example of the amplitudes and phases of the fields is shown in Fig.~\ref{figS:crosstalk_fields}(a) and (b).

Using Eq.~\ref{eq:supp_full_hamiltonian}, we can write the Hamiltonian for the interaction at a given position as
\begin{equation}
	\hat H = \frac{1}{2}\Omega_{\text{DD}} \hat{\sigma}_{+} + \frac{1}{2} \Omega_{\text{SB}} e^{\phi_{\text{grad}} - \phi_{\text{DD}}} \hat{\sigma}_{+} \left( \hat{a} + \hat{a}^{\dagger} \right) + \text{h.c.}\
\end{equation}
where we have neglected off-resonant carrier driving and used the DD phase as a reference.
This choice of reference phase makes it possible to write
\begin{equation}
    \begin{aligned}
	\hat H &= \frac{1}{2}\Omega_{\text{DD}} \hat{\sigma}_{x}\\
    &- \frac{1}{2} \Omega_{\text{SB}} \sin\left(\phi_{\text{grad}} - \phi_{\text{DD}}\right) \hat{\sigma}_{y} \left( \hat{a} + \hat{a}^{\dagger} \right)\\
    &+ \frac{1}{2} \Omega_{\text{SB}} \cos\left(\phi_{\text{grad}} - \phi_{\text{DD}}\right) \hat{\sigma}_{x} \left( \hat{a} + \hat{a}^{\dagger} \right)\\
    \end{aligned}
\end{equation}
making the two parts of the SDF apparent: one which commutes with the DD and one which does not.
A sufficiently strong DD drive would eliminate the effect of the $\hat\sigma_y$ term, following the scaling in Eq.~(\ref{eq:carrier_strength_scaling}).
Combined with an ideal Walsh cancellation scheme, the residual sideband interaction strength, $\Omega_{\text{res}}$, is then given by
\begin{equation}
    \label{eq:effective_interaction}
    \Omega_{\text{res}} = \Omega_{\text{SB}}\cos{(\phi_{\text{grad}} - \phi_{\text{DD}})}\\ .
\end{equation}
To estimate the crosstalk based on this residual sideband interaction strength, we rely on the fact that phase loop-closure -- a key component of high-fidelity two-qubit gates -- is ensured by choosing the correct gate duration and detuning.
Therefore, loop-closure is also enforced on non-addressed ions with a residual sideband interaction.
If an ion experiences an imperfectly cancelled SDF, errors are induced by the phase loop it performs, entangling it with other ions in the chain.
The dominant entanglement occurs with the addressed ions.
We therefore define the parameter $\zeta$ as a measure of the relative strength:
\begin{equation}
    \label{eq:zeta}
    \zeta = \frac{\Omega_{\text{res, non-addressed}}}{\Omega_{\text{res, addressed}}}\\ .
\end{equation}
We estimate the magnitude of this error by considering a two-qubit system of one addressed ion, which is driven by an SDF of amplitude $\Omega_{\text{res, addressed}}$, and a decoupled ion, which experiences a residual SDF of amplitude $\Omega_{\text{res, non-addressed}}$.
The unitary evolution of such a two-ion system can be calculated analytically with the Magnus expansion~\cite{Magnus} and is given by
\begin{equation}
    \label{eq:Magnus}
    U_{\text{MS}} = \text{exp} \left[-i\frac{\pi}{8}\left(\sigma_{x} \otimes \mathds{1} - \zeta \mathds{1} \otimes \sigma_{x} \right)^2\right] .
\end{equation}
This drives the unitary evolution
\begin{equation}
    \label{eq:phase}
\begin{aligned}
    \ket{++}& \rightarrow &e^{-i\frac{\pi}{4}\zeta} &\ket{++} \\
    \ket{+-}& \rightarrow &e^{i\frac{\pi}{4}\zeta} &\ket{+-} \\
    \ket{-+}& \rightarrow &e^{i\frac{\pi}{4}\zeta} &\ket{-+} \\
    \ket{--}& \rightarrow &e^{-i\frac{\pi}{4}\zeta} &\ket{--} .\\
\end{aligned}
\end{equation}
To quantify the crosstalk given by the residual entanglement, we use the impurity of the partial trace of the non-addressed qubit.
Averaging over the three bases ($\hat{\sigma}_{x/y/z}$) gives us the following crosstalk formula
\begin{equation}
    \label{eq:overlap_error}
    \frac{1}{6}\left(1-\cos{\left(\pi\zeta\right)}\right)\ .
\end{equation}
We compute this crosstalk at all positions, as shown in Fig.~\ref{figS:crosstalk_fields}(c), and take the maximum value as the measure of ``maximum crosstalk'' shown in Fig. 5.

\subsection{Gate fidelity}
\label{sec:three_electrode_sims_fidelity}
The ions are addressed when $\phi_\text{DD}=\phi_\text{SB}$.
At these locations, the gradient in decoupling phase $\phi_\text{DD}$ is maximum, and movement of the ion away from the $\phi_\text{DD}=\phi_\text{SB}$ point will suppress the effect of the SDF, potentially affecting two-qubit gate performance.
To simulate this effect, we modify the dynamical decoupling term following
\begin{equation}
	\hat H_{\text{DD}}  = \frac{\hbar}{2} \Omega_{\text{DD}}\left( e^{i \phi_{\text{DD}}} \hat{\sigma}_{+}+ \text{h.c.} \right)+ \frac{\hbar}{2} \frac{\partial \Omega_\text{DD,y}}{\partial u}\hat u\hat \sigma_y\ ,
    \label{eq:H_DD_fidelity_calculations}
\end{equation}
where $\phi_\text{SB}=0$.
Any static shift in the ion position is accounted for by the phase $\phi_{\text{DD}}$.
Dynamical excursions from this equilibrium position (quantum fluctuations, motional heating) are taken into account by including the gradient term weighted by $\partial \Omega_\text{DD,y}/\partial u$.
Here $\Omega_\text{DD,y}$ quantifies the DD field out-of-phase with the sideband field.
As shown in Fig. 3, $\Omega_\text{DD,y}$ is the only field component which varies with position.
We simulate driving a two-qubit gate on an arbitrary pair of ions within a 17 ion crystal as shown in Fig.~\ref{figS:crosstalk_fields}.
To obtain an approximate value for the zero-point fluctuations of motion $u_\text{zpf}$, we consider the axial common-mode frequency required for a 5 $\upmu$m inter-ion spacing, $\omega_m\approx 200$ kHz~\cite{James1998}.
We then simulate driving a 1-loop {M}\o{}lmer-{S}\o{}rensen gate, with a gate time of $250\ \upmu$s (extrapolating from ~\cite{weber2024robust}), and utilising Walsh-15 modulation for the dynamical decoupling.
Despite the zero-point fluctuations $u_\text{zpf}\approx 6$ nm being comparable to the extent of phase variations (see Fig.~\ref{figS:crosstalk_fields}(d)), we find that phase variations across the ground-state wave-packet have little effect on the gate fidelity, inducing an error of $3.8\times 10^{-5}$ (see Fig.~\ref{figS:crosstalk_fields}(f) with $\bar n =0$ at small displacements).
This is because, in the interaction picture, the position-dependent term in Eq.~(\ref{eq:H_DD_fidelity_calculations}) oscillates with frequency $\omega_m$, which is much faster than the gate dynamics.
Additionally, at the interaction zones, the DD field amplitude is minimised as a result of the interference effect producing the desired phase (see Fig.~\ref{figS:crosstalk_fields}(e)).
Similarly, thermal population of the mode has a small effect.
For example, for $\bar n=0.3$, the error increases by $2.7\times 10^{-4}$, whereas without dynamical decoupling, it would increase by $2.4\times 10^{-4}$.
However, gates are severely affected by the average position with respect to the microwave field distribution.
In the example of Fig.~\ref{figS:crosstalk_fields}, a position shift of 5 nm is sufficient to induce an error $>10^{-3}$.
Keeping long term drifts in position below this length scale appears challenging.
In Ref.~\cite{Kranzl2022} for example, the authors report tens of nanometers of axial position drift on the timescale of minutes in a 50 ion chain.
Active stabilisation would thus be necessary for higher-fidelity gates.
This seems achievable, since in previous work~\cite{leu2023} we were able to track the changes in the microwave-induced Rabi-frequency corresponding to $<0.5$ nm with 5 second measurements, using the spatial gradient of the microwave field.

\begin{figure} [t]
    \centering
    \includegraphics[width=0.45\textwidth]{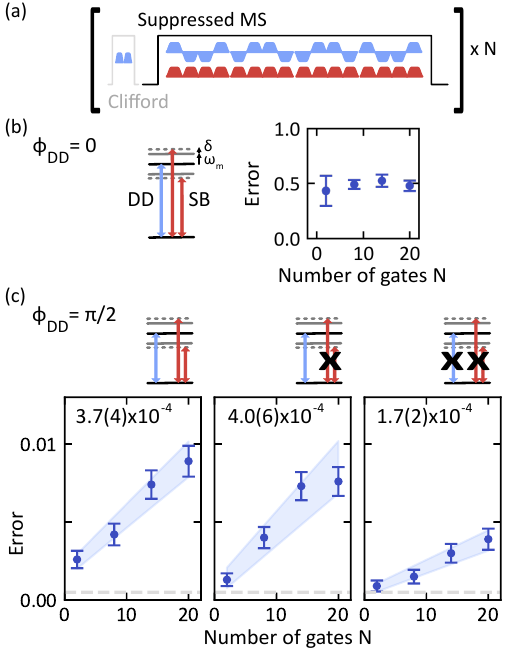}
    \caption{
    \textbf{Randomised benchmarking of the suppression error during a gate.}
    \textbf{(a)} The residual error after suppressing the state-dependent displacement during a gate is experimentally estimated by embedding M\o{}lmer-S\o{}rensen (MS) gate pulses in a single-qubit randomised benchmarking (RB) sequence.
    The effect of the MS interaction is suppressed through a DD drive structured in a Walsh-15 sequence.
    \textbf{(b)} Driving scheme: to closely emulate a (1-loop) MS gate, the detuning from the single-ion motional sidebands $\delta/2\pi=770$ Hz is consistent with both pulse time (1.30 ms) and sideband strength ($\Omega_\text{sb}/2\pi=380$ Hz).
    When the DD is in-phase with the SDF driven by the sidebands ($\phi_\text{car}=0$), the MS pulses entangles the qubit with the rapidly decohering motion, producing the maximum RB sequence error (0.5).
    \textbf{(c)} With the DD out of phase with the SDF ($\phi_\text{car}=\pi/2$), a low error is demonstrated (left panel), on par with the error induced by DD driving alone (center panel) indicating a suppression of the effect of the ion-motion interaction.
    The DD driving error is partially accounted for by leakage and qubit decoherence, measured by replacing the MS pulse by a delay of equal duration (right panel).
    }
    \label{figS:rbm}
\end{figure}

\section{Suppression error}
\label{sec:rbm}

In this section, we provide further details of the single-qubit randomised benchmarking (RB) \cite{Knill2008RBM} measurement, used to estimate the suppression error during a M\o{}lmer-S\o{}rensen (MS) gate \cite{sorenson2000}, which is presented in Fig. 6.
%
%
%
With a sideband strength $\Omega_\text{SB}/2\pi=380$ Hz, an MS gate would be driven by a $1.30$ ms pulse, with sidebands detuned by $\delta/2\pi=770$ Hz.
Driving such a pulse on a single qubit -- with perfectly coherent motion -- would cause $\ket{\pm}$ states to undergo closed loops in phase space and induce no net change in the qubit state.
However, since in this case the motional heating rate (370 quanta/s) is commensurate with the sideband strength, an MS pulse will entangle the qubit with the rapidly decohering motion which in turn decoheres the qubit.
For two representative basis states, the MS pulse (theoretically) transforms the state density matrix following
\begin{equation}
    \begin{aligned}
    \begin{pmatrix}
        1 & 0 \\
        0 & 0
    \end{pmatrix}
    &\rightarrow
    \begin{pmatrix}
        0.61 & 0 \\
        0 & 0.39
    \end{pmatrix}\ ,\\
    \begin{pmatrix}
        1/2 & i/2 \\
        -i/2 & 1/2
    \end{pmatrix}
    &\rightarrow
    \begin{pmatrix}
        1/2 & 0.11\times i \\
        -0.11\times i & 1/2\ .
    \end{pmatrix}
    \end{aligned}
\end{equation}
This will serve as a proxy for the undesired motional and inter-qubit entanglement which would occur in a multi-qubit register as portrayed in Fig. 5.
We aim to gauge how effectively the effect is suppressed upon adding a DD drive with a $\pi/2$ phase shift with respect to the SDF.
To obtain an error averaged over all possible starting states, we embed the MS pulse in a single-qubit randomised benchmarking (RB) sequence \cite{Knill2008RBM} as shown in Fig.~\ref{figS:rbm}(a).
After each single-qubit Clifford gate, we drive an MS pulse and measure the sequence error as a function of the number of Clifford/MS-pulse pairs subjected to the qubit, as typically done in RB.
Given our extremely low Clifford gate error ($1.5\times10^{-6}$) \cite{leu2023}, the sequence error is dominated by the MS-pulses.
Beyond the detuning and pulse timing, there are a few differences with respect to microwave driving in the other presented experiments.
First, we found that the lowest errors are obtained by further increasing the Walsh modulation order from 3 to 15.
Secondly, in order for our control system to implement this more complex pulsing scheme, we had to reduce the pulse ramping duration from $2.4\ \upmu$s to  $1.6\ \upmu$s.
For the same reason, the delay between individual pulses of the Walsh sequence is set to $24\ \upmu$s, and the delay between two Clifford/MS-pulse sequences is $100\ \upmu$s.
In total, an individual Clifford/MS-pulse sequence has a duration of 1.78 ms, which will be relevant when considering decoherence and leakage errors below.
Also, the change in motional mode occupation (due to heating) will be significant for long benchmarking sequences, and so we restrict our experiments to 20 Clifford/MS-pulse pairs.
We restrict our study to a single value of the DD drive strength $\Omega_\text{DD}/2\pi=152$ kHz.

We first run a test measurement with $\phi_\text{DD}=0$, to verify the effectiveness of the RB sequences in revealing errors arising from the MS-pulse (Fig.~\ref{figS:rbm}(b)).
We find that, as expected, the error signal rises to its maximum value (0.5) indicating that the MS-pulses lead to a complete loss of coherence.

\begin{table}[t]
    \begin{tabular}{l|c}
    Error source            & Error ($/ 10^{-4}$ )                 \\ \hline
    \rule{0pt}{2.5ex}Decoherence $T_\text{2}$     & 1.3(2)    \\
    Leakage (from $\ket{0}$)         & 1.5(1)    \\
    Leakage (from $\ket{1}$)        & 0.36(8)    \\
    Detuning                & 0.19     \\
    \hline\hline\rule{0pt}{2.5ex}Expected error         & 3.4(3)                   \\
    Measured error            & 1.7(3)
    \end{tabular}
    \caption{
    \textbf{Time-delay error budget.}
    Expected and measured error budget for the RB sequence where MS pulses are replaced by delays of equal duration (Fig.~\ref{figS:rbm}(c) right panel).
    Mismatch to measurement is attributed to a poor understanding of the leakage error channel, and the difference in time-scales probed by the RB sequence ($\sim$10 ms) and leakage/$T_2$ measurements ($\sim$1 s).
    }
    \label{tab:error_budget}
\end{table}

We then measure the case $\phi_\text{DD}=\pi/2$, obtaining an error per Clifford/MS-pulse pair of $3.7(4)\times 10^{-4}$.
This error is comparable to that obtained with only DD driving ($4.0(6)\times 10^{-4}$), indicating that the limiting factor is not related to motional coupling.
Finally, we measure the error arising if the MS-pulse is replaced by an equally long time delay, revealing that about half of the error, $1.7(2)\times 10^{-4}$, is not related to the microwave driving at all.
These results are shown in Fig.~\ref{figS:rbm}(c).
The expected error from ion-motion coupling is simulated to be between $2.1 \times 10^{-5}$ (when performing a single gate) and $3.0 \times 10^{-4}$ (accounting for heating during 20 sequential gates).
The leading error we were able to identify for DD or sideband driving is the coupling to spectator hyperfine states, detuned by $\sim100$ MHz and featuring low decoherence time scales $\sim 20$ ms.
The error arising from hybridisation of the qubit states with these spectator states and subsequent decoherence is relatively small ($\sim10^{-6}$).
Whilst we cannot explain the origin of the error induced by DD driving, we are reassured by the fact that we cannot identify fundamental limitations to the Walsh sequence, indicating that the dominant error is probably technical.

The error associated with the duration of the pulse, measured when replacing the MS pulse with a delay, is roughly consistent with independently measured error channels (see Table \ref{tab:error_budget}).
Leakage was measured before and after the RB data acquisition by preparing states $\ket{0}$ or $\ket{1}$, waiting for a variable delay and subsequently reading out the qubit state.
The measured error is asymmetric with respect to the starting state, with leakage occurring on time scales $\tau=$ 6.1(6) and 25(5) seconds for starting states $\ket{0}$ and $\ket{1}$ respectively (where $P(\ket{0,1})\approx e^{-t/\tau}$).
Assuming that during the RB sequence $P(\ket{0,1})\sim 1/2$ on average, this corresponds to errors $1.5(1)\times 10^{-4}$ and $0.36(8)\times 10^{-4}$ respectively.
A Ramsey measurement (with varying qubit detuning), performed after the RB sequences, identifies the miscalibration or drift in qubit frequency to be $2\pi\times 1.3$ Hz -- which should only contribute to the error measured without microwave driving, as the DD drive decouples the qubit from detuning offsets.
The impact of this detuning error is estimated through simulating RB sequences to be $0.19\times 10^{-4}$.
A Hahn-echo measurement performed before and after the RB data acquisition reveals a $T_2$-echo time of $4.4(6)$ seconds, consistent with values measured at small time-scales through memory benchmarking~\cite{leu2023}.
This leads to an error of $1.3(2)\times 10^{-4}$.
Mismatch of the total estimated error with the measured value is associated with a poor understanding of the leakage error channel, and the difference in time-scales probed by the RB measurement ($\sim$10 ms) and leakage/$T_2$ measurements ($\sim$1 s).

\end{document}